\documentclass[twocolumn]{article}
\usepackage{usenix}

\usepackage[english]{babel}
\usepackage{blindtext}

\usepackage[title,toc,titletoc,page]{appendix}
\usepackage[ruled,vlined]{algorithm2e}
\usepackage{pdfpages}
\usepackage{array}
\usepackage{amsthm}
\usepackage{xcolor}
\usepackage{optidef}
\usepackage{subfigure}
\usepackage{framed} 
\usepackage{wrapfig}
\newcolumntype{L}{>{$}l<{$}} 
\newcolumntype{R}{>{$}r<{$}} 
\usepackage{enumitem}

\newtheorem{definition}{Definition}

\newtheorem{proposition}{Proposition}

\newcommand{\system}{\emph{Carbide}\xspace}
\newcommand{\impl}{\emph{MultiJet}\xspace}
\newcommand{\DistVeri}{\emph{CPCheck}\xspace}
\newcommand{\FSD}{\emph{FSD}\xspace}
\newcommand{\ReqLang}{\emph{ReqLang}\xspace}
\newcommand{\CPSpec}{\emph{CPSpec}\xspace}
\newcommand{\para}[1]{\noindent {\bf #1}}

\newcommand{\reducespace}{\vspace{-0.1in}}

\newcommand{\dnet}{DV-Network\xspace}
\newcommand{\ie}{\emph{i.e.}}
\newcommand{\eg}{\emph{e.g.}}

\long\def\qiao#1{{\color{orange}{\bf Qiao: }{\small #1}}}

\long\def\review#1{{\color{red}{\bf Review: }{\small #1}}}

\begin{document}
\pagenumbering{gobble}
\onecolumn
\newpage
\clearpage
\thispagestyle{empty}
\includepdf[pages=-,pagecommand={},width=1\textwidth]{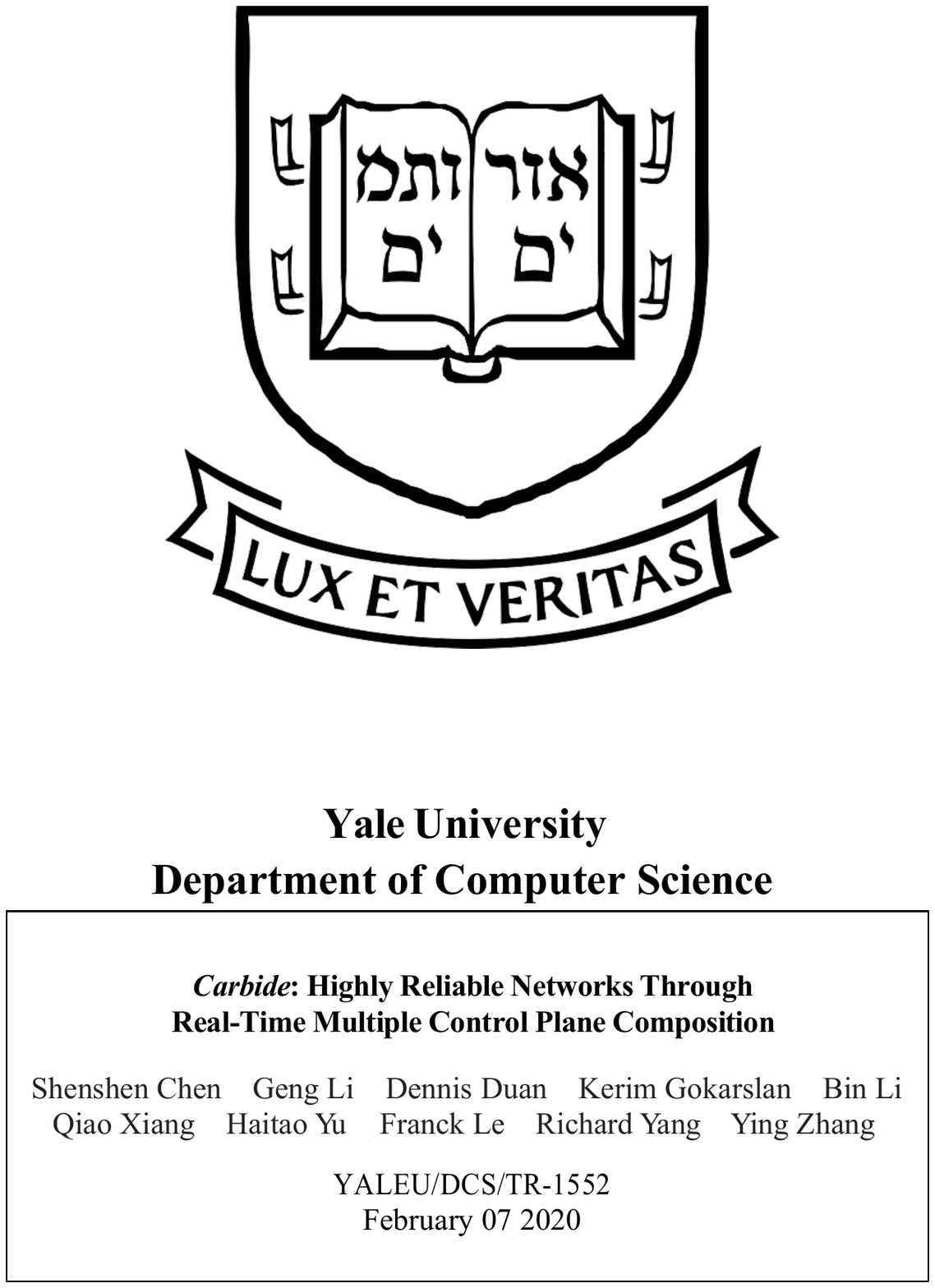}

\twocolumn
\newpage
\clearpage
\pagenumbering{arabic}
\begin{abstract}
	Achieving highly reliable networks is essential for network operators to ensure
proper packet delivery in the event of software errors or hardware failures.
Networks must ensure reachability and \textit{routing correctness},
such as subnet isolation and waypoint traversal. Existing work in network
verification relies on centralized computation at the cost of fault tolerance,
while other approaches either build an over-engineered, complex control plane, or compose multiple control planes without providing any
guarantee on correctness. This paper presents \system{}, a novel system to
achieve high reliability in networks through distributed verification and multiple control plane composition. The core of \system is a \textit{simple, generic, efficient distributed verification framework} that transforms a generic network verification problem to a reachability verification problem on a directed acyclic graph (DAG), and solves the latter via an efficient distributed verification protocol (DV-protocol).  Equipped with verification results, \system{} allows the systematic composition of multiple control planes and realization of operator-specified consistency.
\system is fully implemented. Extensive experiments show that (1) \system reduces downtime by 43\% over the most reliable individual underlying control plane,
while enforcing correctness requirements on all traffic; and (2) by systematically decomposing computation to devices and pruning unnecessary messaging between devices during verification, \system{} scales to a production data center network.
\end{abstract}

\footnote{This technical report includes Dennis Duan's undergraduate thesis.}

\vspace{-1em}
\section{Introduction}
The expectation for network availability is increasingly demanding. For example, Google increased its service level objectives (SLOs) from 99\% availability in 2013 to 99.99\% in 2018~\cite{hong2018b4, jain2013b4}.
This is because business-critical applications are increasingly reliant on networks, and the cost of infrastructure downtime can now reach \$7 million per hour~\cite{outage}. 

Given its importance, substantial efforts have been devoted to increasing network reliability~\cite{beckett2017general,gember2017automatically,handigol2014know,kazemian2013real,kazemian2012header,khurshid2013veriflow,zeng2012automatic,kvalbein2006fast,tilmans2014igp, vissicchio2015co, jayaraman2019validating}. 
A major advance to improve reliability is to use network verification. 
Many verification methods~\cite{kazemian2012header, khurshid2013veriflow, kazemian2013real, yang2015real, horn2017delta} model all possible forwarding behaviors and compute possible violations of network requirements. Other approaches convert network state and requirements into systems of boolean constraints and utilize SAT or SMT solvers to compute correctness~\cite{beckett2017general, gember2017automatically}. Complementary to verification, active testing~\cite{handigol2014know, zeng2012automatic} or control plane emulation~\cite{liu2017crystalnet, lopes2019fast} are used to catch bugs after the deployment. On the other hand, network synthesis~\cite{propane,merlin,contra,el2018netcomplete} tries to avoid problems by systematically generating configurations. 



Despite considerable progress, existing approaches still suffer major limitations. 
First, most verification tools rely on a centralized verification server to collect snapshots of the network states or configuration files from all of the network devices, which requires reliable connections between the server and network devices, resulting in a bottleneck and a single point of failure. Azure~\cite{jayaraman2019validating} proposed to have devices verify its forwarding behaviors using local contracts. However, they only support  verifications of the shortest path reachability and fault tolerance property. Secondly, the attempt of building a single, infallible control plane often results in over-engineered, complex control plane.  
Third, existing work in multiple control plane composition supports only a limited number of properties (\eg, reachability, domain backup) and fail to provide generic routing correctness guarantees (\eg, waypoint routing, subnet isolation)~\cite{le2010theory}.

This paper systematically investigates and tackles the aforementioned limitations of existing approaches to improve network reliability, and presents \system{}, a novel system to
achieve high reliability in networks through distributed verification and multiple control plane composition. 

Specifically, the core of \system is \DistVeri, a \textit{simple, generic, efficient distributed verification framework} that lets ingress devices verify which packet space can be forwarded by a given CP without violating requirements specified by the operators. \DistVeri has two key insights:
(1) A generic verification problem (\eg, reachability, loop-free, waypoint and fault-tolerance) on a generic network can be transformed to a reachability verification problem on a directed acyclic graph (DAG), and (2) the latter can be solved via a novel, efficient distributed verification protocol (DV-protocol). By systematically decomposing computation to each device and pruning unnecessary messaging between devices, \DistVeri scales to a production data center network. Rigorous analysis proves the convergence and correctness of \DistVeri. It also shows that the previous published Azure local verification~\cite{jayaraman2019validating} can be easily supported as a specialized case of verifying shortest-path reachability and fault-tolerance requirement in \DistVeri.

Next, to allow systematic multiple control plane composition, \system{} provides (1) \textit{CPSpec}, a flexible grammar that allows operators to specify correctness requirements for each CP on different packet space, the preference relation between CPs, and the desired consistency model, and (2) \textit{CPComposer}, a module that uses the verification results of \DistVeri, select different CPs to use for different packet spaces, and schedules the hot-swapping of the corresponding data plane to guarantee the consistency requirement specified in \textit{CPSpec}.  
Moreover, by constructing virtual CPs to pre-verified alternative next-hops, tunnels and the mix of different CPs, \system{} provides the verified fast-reroute (V-FRR) capability to the network, which substantially reduces the network downtime while guaranteeing routing correctness.

 A switch OS software suite called \emph{Multijet} is implemented to deploy \system on real white-box switches. Extensive experiments show that (1) \system reduces downtime by over an order of magnitude compared to SDN, and by up to 43\% when compared to OSPF. Even in the presence of network partitions, \system correctly enforces network requirements (\eg, on security, waypoint) on all packets, and (2) by systematically decomposing computation to devices and pruning unnecessary messaging between devices during verification, \system{} scales to a production data center network with little overhead. This work does not raise any ethical issues.

 
\vspace{-1em}
\section{Motivation}\label{sec:motivation}


\para{Problems with centralized verification.} Always ensuring the network functions as desired, even in face of failure, is challenging. Existing verification~\cite{ kazemian2012header, khurshid2013veriflow, kazemian2013real, yang2015real, horn2017delta} fails to meet this requirement. Although differing in details, these work employ a common centralized architecture: a centralized server is used to collect data from each network device and verify the invariant compliance. This architecture has two major pitfalls to satisfy the "always correct" requirement. First, it heavily relies on the network connection between the verification server and the devices. The network connection, however, is possibly congested or broken during outage, especially catastrophic failure. Second, the server becomes the bottleneck and single point of failure. Fundamentally, it violates the \textit{fate sharing principle}~\cite{clark1988design}: the network control and data path should share the same fate, they either fail together, or not at all. 

The fate sharing principle naturally motivates the distributed verification design. Verification messages should flow on the same path as traffic. The sender of the path shares the same fate as its traffic, thus, the source/ingress device should be responsible for verification rather than an off-path centralized server. A naive approach is for all devices to send relevant information (\eg, FIB entries) to the source so that it can run various verification on its own traffic. Yet, flooding FIB to all devices certainly is not scalable nor necessary. 

Our key idea is to distribute the verification and systematically prune the messages to eliminate most of the unnecessary communications. Each device can make local verification and propagates messages to source only when its local  verification results change.
The communication is limited to the small set of switches according to the data plane path and requirements. 

\para{Problems with a single control plane.} While distributed verification provides better correctness guarantee during failure, high availability is still not satisfied because of a single control plane. 
One may argue that a single control plane can survive failure by setting up multi-paths or fast reroute capabilities. However, it often results in an over-engineered, complex control plane. More importantly, these data plane bandaids cannot handle control plane bugs and misconfigurations.

High availability requirements call for multiple control plane coexistence. Internet is a great example of running multiple control plane to tolerate failure, rather than relying on a single ``never-failed" control plane. At the macro level, each Autonomous System controls its own network independently so that the failure's impact can be constrained within the domain. Within a single network, it is not uncommon to have multiple control planes in production. For example, multiple controllers are used to control different planes of a backbone~\cite{ebb,jain2013b4}. At the micro level, multiple routing protocols (\eg, OpenR, BGP) can run together on each switch~\cite{fboss}. These designs use multi-control plane to ensure high availability. 

Naively running multiple control planes simultaneously does not work since different control plane may make conflicting decisions and result in violation of network policy. Thus, we use distributed verification results to compose control planes to guarantee correctness and resilience.

\section{\system Overview}\label{sec:overview}
\begin{figure}[t]
\setlength{\abovecaptionskip}{0.1cm}
\setlength{\belowcaptionskip}{-0.cm}
\centering
\includegraphics[width=\columnwidth]{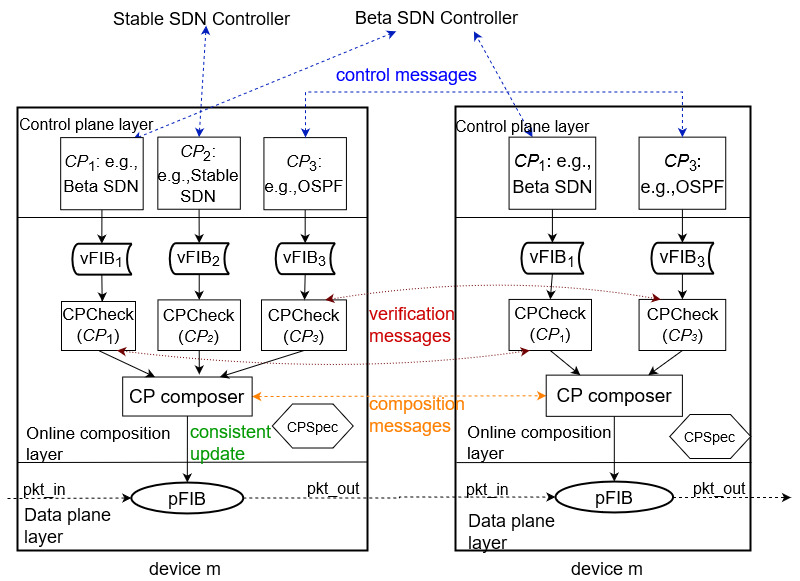}
\caption{\small{\system architecture.}}
\label{fig:arch}
\reducespace
\vspace{-1em}
\end{figure}


\system{} is a thin layer between CP and switches (Figure~\ref{fig:arch}). Specifically, each network device runs (1) a control plane layer, (2) a novel online composition layer, and (3) a unified data plane layer.





\subsection{Control Plane Layer} 
The control plane layer consists of a set of control plane (CP) instances $\mathcal{CP} = \{CP_i\}_i$, $i=1, 2, \ldots$, running in parallel. A control plane instance may be centralized (\eg, SDN) or distributed (\eg, OSPF and BGP). For example, a device may run three control plane instances: $CP_1$ as an SDN control plane receiving OpenFlow messages from a new release of an SDN controller, $CP_2$ as an SDN control plane receiving OpenFlow messages from  a stable release of an SDN controller, and $CP_3$ as a traditional distributed routing protocol such as OSPF.
\system treats every control plane instance as a black-box 
and only depends on the output (\eg, forwarding information base) of each control plane instance. This design decision allows \system to use any existing implementation (open source or commercial) for each control plane instance.

\subsection{Online Composition Layer} 
This novel layer dynamically composes the information from the control plane instances to satisfy the requirements. It contains four components.


    \para{\CPSpec}. A global specification allows operators to specify correctness requirements for each CP of different traffic types, and the preference order between control plane instances. 
    Specifically, a \CPSpec is specified as a tuple of ($ps$, $\{(req_i, CP_i)\}_i$, $option$). $ps$ is the packet space of interests expressed as a predicate of packet headers. Given a $ps$, a set of packet ingestion points (ingress of packets in $ps$) will be identified. A CP composer will be started at each such source. 
    
    $\{(req_i, CP_i)\}_i$ is a sequence of (requirement, CP) pairs in descending order of CP preference. Because different CPs may have different desired behavior, \CPSpec allows operators to specify \emph{different} requirements for each CP. As such, if a CP cannot satisfy its desired requirement, it will not be used. 
    A requirement $req_i$ is expressed as a predicate of regular expressions, which describes a set of paths in the network (\eg, reachability, waypoint, and loop-freeness). Section~\ref{sec:dv-grammar} gives the details of the requirement grammar. 
    
    $option$ allows an operator to specify the consistency model to enforce. \system{} guarantees the eventual consistency by default, and can achieve a stronger consistency (\eg, transient-loop-free consistency) with a trade-off on performance. (\S~\ref{sec:consistency}) 
    

       
    \para{vFIB}: Each switch running a CP is associated with a virtual forwarding information base that collects and stores the forwarding information computed by the CP (\eg, FIB and ACL). vFIB does not need an internal forwarding state of a device (\eg, counter). For the same packet, a CP may have multiple next-hops in its forwarding information (\eg, load balancing, robustness, and multicast). vFIB puts all these next-hops as a group action, and does not need the underlying hardware realization.  

    \para{\DistVeri (Section~\ref{sec:dv})}: Each device running $CP_i$ is associated with \DistVeri, a simple, efficient distributed verification component.  The goal of \DistVeri is to let each source identified by \CPSpec to verify which packets can be forwarded by $CP_i$ without violating related requirements specified by the \CPSpec. The core idea of \DistVeri consists of (1) transformation of a generic requirement verification on a generic network to a reachability verification on a DAG, and (2) a novel, efficient DV-protocol to verify reachability on the DAG. The root of the DAG is the source (see Section 4.4).

    
    \para{CP composer (Section~\ref{sec:composition})}:  The CP composer takes as inputs the forwarding rules from the vFIBs, the verification results from the \DistVeri modules, and the consistency model to compute the CP assignments of packets to enforce at the data plane layer. 
    In the general case, when traffic in the same packet space enters the network from multiple ingress points, these devices run a consensus protocol to let one decide and announce the CP assignments. See Section~\ref{sec:dv-transformer} for examples of the complete workflow.

  

\para{Resource control.} To improve performance, \system{} develops a resource control component on each device to allocate shared resources (\eg, CPU, memory and bandwidth) among different running processes (\eg, control plane instances, \DistVeri modules, CP composer). Specifically, this component proactively limits the resource consumption of each process (\eg, CPU and bandwidth usage), to prevent a process using up all of the resources on a device and starving all other process. It also adaptively adjusts the resources allocated to different processes in response to the behaviors of CPs. For example, a CP may oscillate between various routes, \system develops a BGP-inspired damping mechanism to control the resources allocated to the corresponding \DistVeri process, to go from passive verification to more active filtering.

\vspace{-4mm}
\subsection{Data Plane Layer} 
Despite multiple CP instances running, each device in \system has a single unified data plane that processes data packets at the line rate based on rules installed by the CP composer in the device's physical FIB (pFIB). Conceptually, the pFIB at the data plane is comprised of the rules from different CPs. A challenge is that different CPs may overlap and conflict, but a packet should only match the rules from one CP. To this end, \system leverages the multi-table structure in commodity switches (\eg, ~\cite{openflow1-4, trident4}) and stores forwarding rules from different CPs in different tables. The CP composer generates an extra CP selection table to safely direct the traffic in the data plane using the corresponding rules. 


\section{\DistVeri: A Distributed Verification Framework} \label{sec:dv}
\DistVeri is a core component of \system{}. At the same time, it
can be a generic distributed verification technique used separately in other settings (\eg, trouble shooting~\cite{handigol2014know}).
Assume a stable data plane, \DistVeri\ allows ingress devices to verify which packet space can be forwarded without violating requirements specified by the operators. 
The key insight of \DistVeri\ is to transform a generic verification problem into a simple \dnet\ problem, and solve it with a novel, efficient DV-protocol.
In the following, we first introduce the data plane model, the language to define requirements and present our detailed algorithm. 

\begin{figure}[t]
\setlength{\abovecaptionskip}{0.1cm}
\setlength{\belowcaptionskip}{-0.cm}
\centering
\includegraphics[width=0.5\columnwidth]{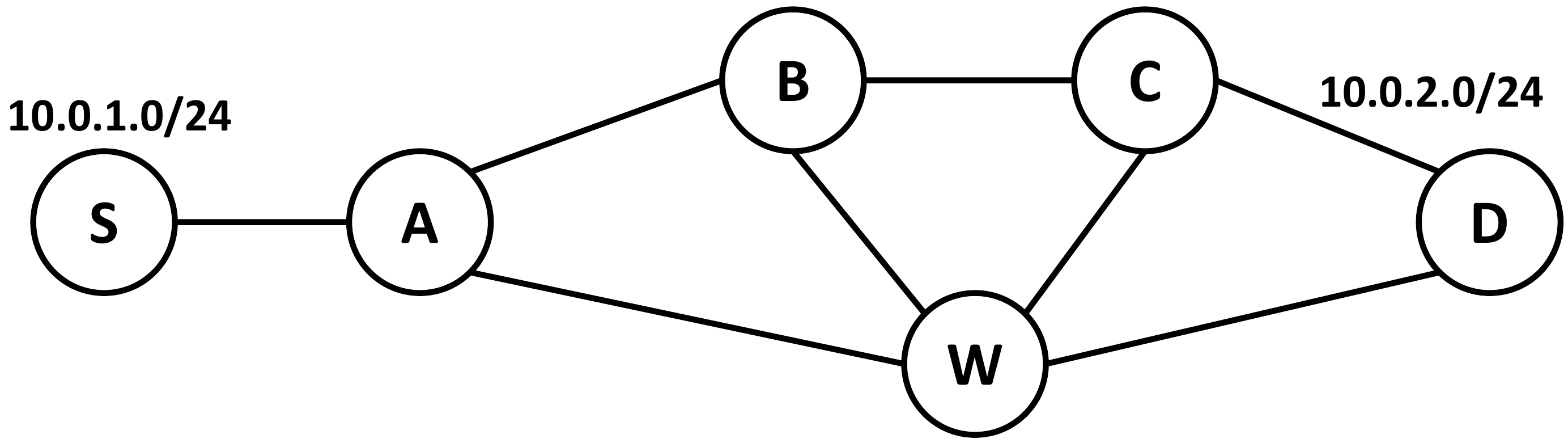}
\caption{\small{An example network.}}
\label{fig:dv-example-topology}
\reducespace
\vspace{-1em}
\end{figure}

\vspace{-2mm}
\subsection{Data Plane Model}\label{sec:dv-dpmodel}

\DistVeri\ uses a generic data plane model that abstracts different FIB heterogeneity, \eg, destination-based FIB, OpenFlow table. 
Each entry in the table maps a disjoint packet space~\cite{kazemian2013real} to an action. The action includes modification of packet header as well as a \emph{group} of next-hops~\cite{jayaraman2019validating}. 
If an action sends a matched packet to
all next-hops in this group, it is annotated with a keyword $ALL$. If it sends to only one next-hop of the group, it is annotated with a
keyword $ANY$. 

This representation allows modeling of different forwarding behaviors at a
network device. For example, drop is modeled by an empty next-hop group.
Packet encapsulation and decapsulation are modeled by modification
functions. Multicast is modeled by an  
$ALL$ next-hop group. Anycast, load-balancing and
fault-tolerance~\cite{jayaraman2019validating} is modeled by an $ANY$ next-hop
group. Given an action, however, \DistVeri is oblivious of its hardware
realization (\eg, how to select one next-hop from an $ANY$ next-hop group). 

Another important concept at the data plane is path segments. The global data plane $DP_i$ for a given control plane $CP_i$ is composed of a set of FIBs, $\{FIB_i^j\}_j$, \ie, the FIBs of all network devices $j$. Each FIB  indicating next-hop constructs 1-hop segments. Together, they construct path segments starting from source devices or other devices.
A path segments is a sequence of network devices. When a device appears twice, it forms a loop.

\vspace{-2mm}
\subsection{Verification Requirements Specification} \label{sec:dv-grammar}

\begin{figure}
\ttfamily
\footnotesize
\begin{tabular}{r L  l}
       h & \in & packet headers
    \\ f & \in & header fields
    \\ v & \in & header field values
    \\ label & \in & device labels 
    \\ packet-space & := &  (h.f [=, $\neq$] v)
    \\ & | & (packet-space [$\cap$, $\cup$] packet-space)
	\\ RegExr & := & ...  extended with: `[' (\^{})label `]'
    \\ path-set & := & RegExr
    \\ & | & (path-set [$\cap$, $\cup$] path-set)
    \\ sources & := & `[' (\^{})label `]'
    \\ requirement & := & (sources:) packet-space $\rightarrow$ path-set
\end{tabular}
	\caption{\small{A simplified grammar of \DistVeri \ReqLang.}}\label{fig:dv-grammar}
\reducespace
\vspace{-1em}
\end{figure}

Next, we define a grammar to specify global requirements. A wide range of common, important verification requirements (\eg, reachability, waypoint, loop free and fault tolerance) can be expressed using grammar shown in Figure~\ref{fig:dv-grammar}. At the high level, a requirement is specified as a \textit{path-set} for a pair of \textit{sources} and \textit{packet-space}. The \textit{sources} optionally refers to devices responsible to verify the requirement. The \textit{packet-space} is defined by predicates of header fields, which is passed by \CPSpec. A pair of \textit{sources} and \textit{packet-space} specifies packets within \textit{packet-space} originated from \textit{sources}. The \textit{path-set} is defined by (dis)conjunction of regular expressions (Figure~\ref{fig:dv-grammar}). 

By default, \DistVeri\ assumes loop-free and only considers correctness requirement. The correctness is defined as path segments starting from source should belong to the  \textit{path-set}. 
For example, the data plane must construct path segments ending with the destinations if the \textit{path-set} consists of paths reachable to the destination. More general requirements are tackled in Section~\ref{sec:dv-extension}.

The grammar of regular expression is mostly standard, but includes $[label]$ as a syntax sugar. It refers to any device with the $label$ and $[$\^{}$label]$ refers to any device without that $label$. A device can have multiple labels (\eg, identifier, ip address, functionality) and can share a label with others.

\para{Example}. Consider a network in Figure~\ref{fig:dv-example-topology}. $S$, $A$ and so on are the identifier labels of network devices. $S$ and $D$ each has an IP label 10.0.1.0/24 and 10.0.2.0/24, respectively, denoting to the subnet each one is connected to. The following requirements assigned to source device $S$. They require the traffic from source IP 10.0.1.0/24 to destination IP 10.0.2.0/24 is always delivered (\ie, reachability) after passing $W$ (\ie, waypoint), without any loop (\ie, loop-free). 

{
\begin{enumerate}
    \item packet-space = (srcIp = 10.0.1.0/24) $\cap$ (dstIp = 10.0.2.0/24)
    \item loop-free = $\bigcap_{\text{device } x}$ (([\^{}$x$]*) $\cup$ ([\^{}$x$]*[$x$][\^{}$x$]*))
    \item reachability = [10.0.1.0/24].*[10.0.1.0/24]
    \item waypoint = .*[W].*
    \item path-set = reachability $\cap$ waypoint $\cap$ loop-free
    \item requirement = ($[S]$: packet-space) $\rightarrow$ path-set
\end{enumerate}
}

Specifically, the first line defines \textit{packet-space} to be the space of packets with source IP 10.0.1.0/24 and destination IP 10.0.2.0/24. Next, it uses regular expressions to define the set of paths satisfying reachability, waypoint and loop-free requirements, respectively. 
In line 5, the \textit{path-set} takes intersection of those sets, which means satisfying all the requirements. 
At the end, the verification requirement is assigned to the source device with identifier label $S$. 
\subsection{Distributed Verification for \dnet} \label{sec:dv-protocol}
Given the preceding requirements, a straightforward approach is to have every node send each FIB entry to the verifier (source in this case). This, however, is unnecessary and undesirable, as it leads to less distributed load. A key novelty of \DistVeri\ is its ability to distribute the verification.

\para{\dnet\ verification problem.}
Instead of starting with generic network verification, we first study the \textit{\dnet\ (distributed verification network)} verification problem. The problem may look special, but it actually is not. We will show in Section~\ref{sec:dv-transformer} how to convert a generic network verification problem to this problem. But for now, we focus on this problem.

Specifically, a \dnet\ is a directed-acyclic-graph (DAG) with only a source and only destinations as sinks. Each node in the network has an FIB to look up the next hop for each packet.  The FIB may return an outgoing edge in the DAG from the node or off-path. The network is verified for a destination if the FIBs construct a path from the source to the destination. 

\begin{figure}[t]
\centering
\includegraphics[width=0.6\columnwidth]{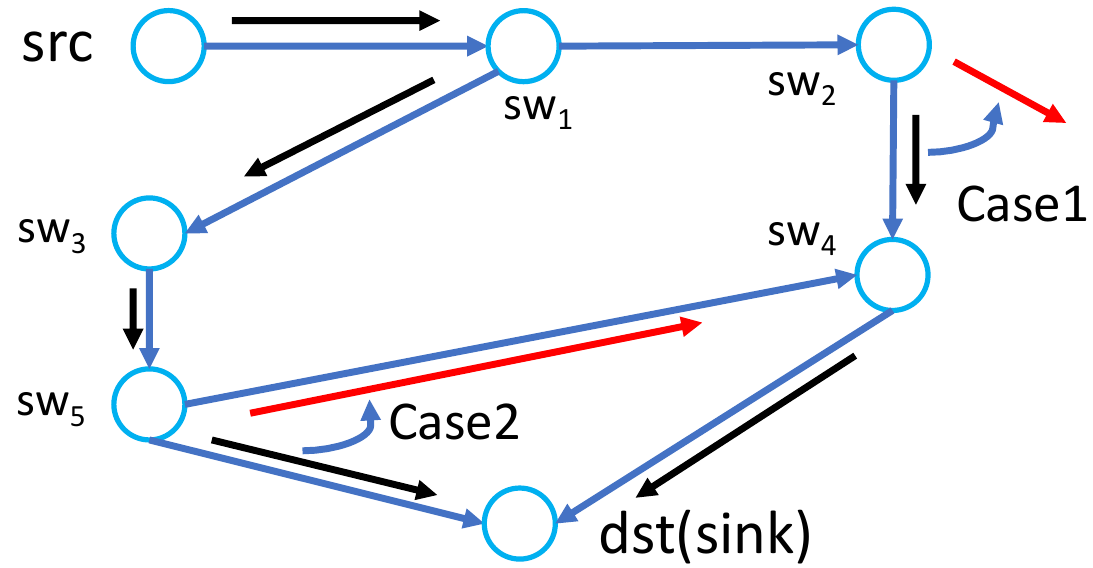}
\caption{\small{Two cases of update for \dnet.}}
\label{fig:dv-dnet}
\reducespace
\vspace{-1em}
\end{figure}

\para{Intuition-building example.}
To build intuition, consider an example \dnet\ verification problem shown in Figure~\ref{fig:dv-dnet}. Note that the example has only a single destination and hence we just consider this destination. Observe that at a given state, the FIBs of the nodes in the \dnet\ form a set of path segments. For example, there exists a path segment $src \rightarrow sw_1 \rightarrow sw_3 \rightarrow sw_5 \rightarrow dst$. The FIBs also set up a 
path segment from $sw_2\rightarrow sw_4 \rightarrow dst$. 



One might think that to verify $dst$, the source needs to know each FIB state at {\em all} time. This, however, is unnecessary. To appreciate it, consider some changes to FIBs and observe whether the source needs to know about the change.

\begin{itemize}[leftmargin=*]
    \item Case1: $sw_2$ updates its nexthop to somewhere outside the network (shown in red line). From a purely local view, it is a violation so $sw_2$ attempts to report it to $src$. However, from a global view, this change has no effect on the path segment chosen by $src$ and hence does not need to be reported. To approximate the global view, $sw_2$ should report the change to its upstream $sw_1$. Since $sw_1$ is not using $sw_2$ as next hop, $sw_1$ can locally decide that there is no need to propagate this change further. This example shows one insight of eliminating unnecessary propagation: if the node is not on the current path from $src$ to $dst$, its change does not need to be propagated.
    \item Case2: $sw_5$ updates its nexthop to point to $sw_4$. This change does not need to report to $src$ either, because $sw_5$'s new change still satisfies the requirement. This illustrates our second insight of reducing propagation: only propagate if the local verification result changes. 
\end{itemize}

\para{Verification function.} To ground the above intuitive examples, formally, we introduce a Boolean verification function $d(x)$: whether the requirement is satisfied at switch $x$, from the global view. 
One can
see that the goal of verification is to compute verification function $d(src)$, which 
can be computed cumulatively from the $d(x)$ along the path.
Specifically, denote the next hop of switch  $x$ (searched using its FIB) as $n(x)$. If the next hop is off network, it is a special switch $nil$. We have:
\begin{equation} \small
d(x) = d(n(x)).
\label{eqn:main}
\end{equation}
As the boundary, $d(dst) = True$ and $d(nil) = False$. 

One can compute Equation \eqref{eqn:main} using a distributed algorithm, and a natural design is a push based design:
each switch $x$ owns its value of $d(x)$ and pushes the value upstream. One can see that it is 
sufficient for $x$ to follow a simple local update rule:
\emph{switch $x$ will propagate $d(x)$ only if its value changes}. 
Let us go back to our example to illustrate the benefits of this rule. For Case1, initially, we have $d(src) = d(sw1) = d(sw3) = d(sw5) = d(dst) = True$. Upon $sw_2$ making the change of its nexthop, $d(sw2)=False$. $sw_2$ propagates its change to $sw_1$, however, because $d(sw_1)$'s computation does not rely on $sw_2$, $d(sw_1)$ remains $True$. Then $sw_1$ will not propagate the change further. Similarly, for Case2, $d(sw_5)=True$ before and after the change, so $sw_5$ will not even trigger the propagation. 
Consider a general case of \dnet\ of an n-by-n grid, with the source at the upper left, destination at the lower right, and each node can go only right or down inside the grid. Assume that each node has computed a next hop (right or down) to the destination. Consider any flip between right and down, there will be no message generated.



At this point, one might make an observation that the algorithm has a structure similar to traditional distance vector routing. Observing this similarity can help understand our design. One may consider verification as computing distances in a domain with only two values: finite (reachable or verified) and infinite (unreachable or unverified). Note that traditional distance vector routing can have churns during convergence. A \dnet, on the other hand, is a DAG and the propagation follows the reverse links of the DAG; hence there are no loops, and hence no churns. 

\para{Single destination to packet space.}
The above cases consider an individual destination. Now consider a \dnet\ with multiple destinations. It is more efficient to consider all destinations together than consider them one by one. With multiple destinations, the FIB at node $x$ may choose different nexthops for different destinations (or more fine-grained packet space determined by both destination and port number). We extend $n(x)$ function to $n_p(x)$, representing the nexthop for packet $p$ in node $x$'s FIB. As a result, $d_p(x)$ is extended to consider different destination space $p$.


\vspace{-0.5em}
\begin{equation}\label{eqn:packet_dv}
d_p(x) = d_p(n_p(x)).  \small
\end{equation}

\para{From equation to DV protocol.}
While Equation (\ref{eqn:packet_dv}) is a compact model, it requires efficient implementation. 
Define $x.H = \{p \; | \; d_p(x) = False\}$ as a header-space~\cite{kazemian2012header},
compact~\cite{kazemian2013real, yang2017scalable} representation for $d_p(x)$. 
Protocol~\ref{alg:dv} gives the basic protocol using header-space operations ($\cap, \cup, -$), for immutable packet headers. One can extend the basic protocol to handle header modification (\eg, fields modification, encapsulation).


\begin{algorithm}[h] \footnotesize
    \SetAlgoLined
    \KwData{Incoming message $y.\Delta H$ indicating $\forall p \in y.\Delta H, d_p(y)$ changes}
    \KwResult{$d_p(x)$ (stored as $x.H$)}
    
    \tcc{Apply negation to local record of $d_p(y)$}
    $y.H = (y.H - y.\Delta H) \cup (y.\Delta H - y.H)$\;
    Recompute $x.H$ by local records $\{y.H\}_y$\;
    \If{$x.H \neq x.oldH$} {
        Propagate the change to upstream\;
    }
\caption{(DV protocol) at node $x$.}
\label{alg:dv}
\end{algorithm}

\para{Convergence and correctness}. We give the following proposition for the convergence and correctness of our
DV protocol:

\begin{proposition}[Convergence and Correctness of Protocol~\ref{alg:dv}]
Assume that (1) \dnet\ is at a stable state, (2) each device executes Protocol~\ref{alg:dv}, and (3) all messages are delivered reliably. Protocol~\ref{alg:dv} always converges. After it converges, for any packet $p$, $d_p(src)$ represents whether $p$ can be delivered to its destination in \dnet.
\end{proposition}



\vspace{-1em}
\begin{proof} \textit{Sketch}.
The convergence of Protocol~\ref{alg:dv} is guaranteed by the fact that \dnet is a DAG. 
For its correctness, first consider the case that Protocol~\ref{alg:dv} executes in a blocking way, \ie, each node $x$ waits till all its downstream neighbors $y$ to send $d_p(y)$, and then computes $d_p(x)$.  With the Equation (\ref{eqn:packet_dv}), it is easy to see that the protocol converges to a state where $d_p(src)$ correctly indicates whether $p$ can reach destinations. Now consider an async model with out-of-order messages, the key observation of correctness is a message of $d_p(y)$ always indicates $d_p(y)$ value has changed. So that $x$ can apply negation to the local record of  $d_p(y)$ value. Therefore, the computation is not affected by the order of messages.
\end{proof}

\vspace{-1em}
\subsection{Transform Generic Networks \& Requirements to DV-Networks} \label{sec:dv-transformer}
\para{Basic issue.} The \dnet\ verification problem appears simple because each simple path in the DAG from the source to the destination is a legitimate path (\ie, can be verified). Such a simple path, however, may not satisfy more general requirements. For example, given a waypoint requirement, a simple path from the source to the destination may not go through the given waypoint and hence is not legitimate; given a shortest path requirement, a simple path may be longer than the shortest path, and hence is not legitimate.

\begin{figure}[t]
\centering
\includegraphics[width=1\columnwidth]{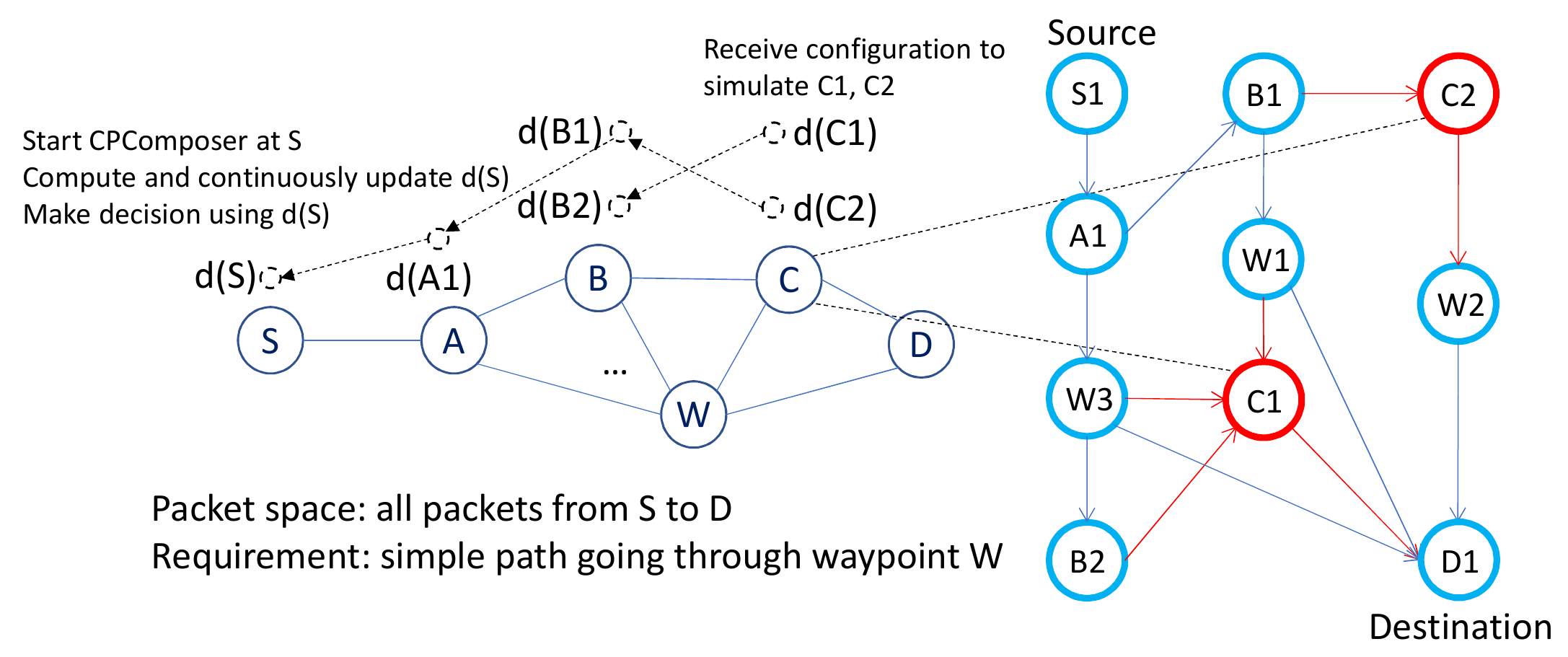}
\caption{\small{\dnet for general topology and waypoint requirement}.}
\label{fig:dv-ddt-mapping}
\reducespace
\end{figure}

\para{Automata requirements to \dnet\ verification.} 
As a large gap as one might think between simple \dnet\ verification and generic verification, one class of general requirements which can be converted to \dnet\ verification is those which can be handled
by the systematic work based on product graphs of automata (\eg, ~\cite{propane}). In particular, Figure~\ref{fig:dv-ddt-mapping} (left) shows a general topology with requirements including reachability, loop-free, and way-point. The right side shows the constructed \dnet. Note that the nodes in \dnet\ are not just 1-1 mapped to nodes in the physical topology. For example, $C$ is split to $C_1$ and $C_2$ because packets can reach $C$ via either $B$ or $W$. Similarly, $W$ is split to three nodes. The construction is essentially based on a {\em path prefix} technique and we refer readers to existing work (\eg, ~\cite{propane}), as the key goal of our work is distributed verification and we leverage existing work as much as possible. In such a setting, a device in the real topology will simulate multiple nodes in \dnet, with a distributed protocol. We show the workflow of verification triggered at $s$ and each node separately compute their $d$ values. Upon failures, $d$ changes propagates back to $s$.

\begin{figure}[t]
\centering
\includegraphics[width=1\columnwidth]{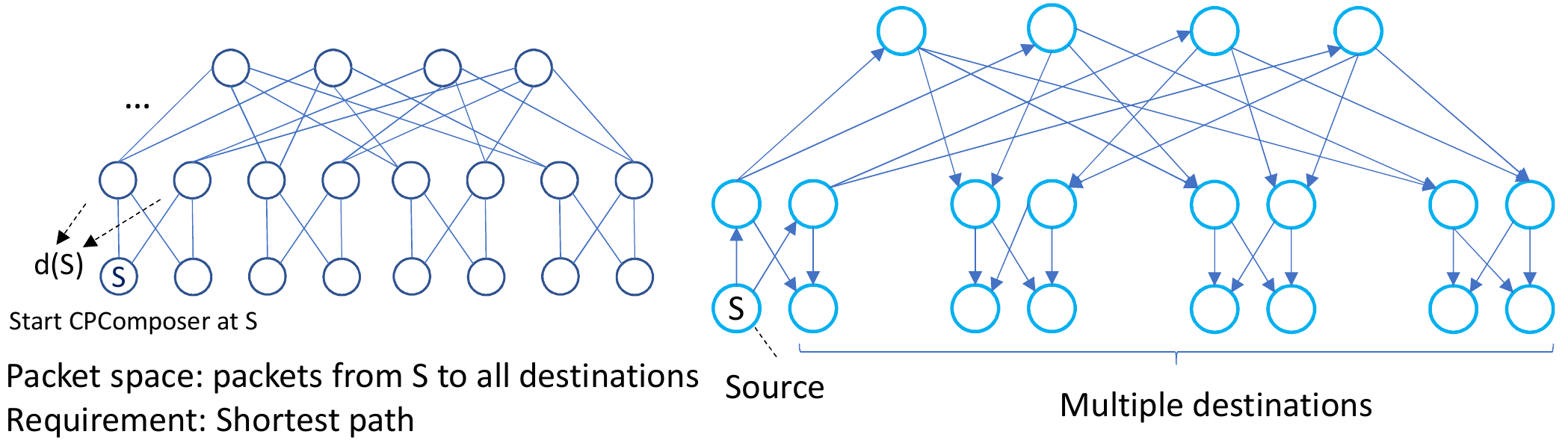}
\caption{\small{\dnet for DC topology and shortest path requirement.}}
\label{fig:dv-1srcNdst}
\reducespace
\vspace{-1em}
\end{figure}
\para{Optimization verification using duals.} Product graphs are not the only way for transformation. 
Optimizations which can be verified by local dual variables may also be converted. Consider (fixed-topology) shortest path, which is an optimization problem which can be used to generate local verification, allowing one to transform problems to \dnet\ verification. Figure~\ref{fig:dv-1srcNdst} (left) shows a DC topology and the requirement of using shortest paths. The source and destinations are marked. The results of shortest-path optimization can be verified using dual variables (distance to the destination in a fixed topology). The right side of shows the \dnet from source $S$ to the set of destinations.
\vspace{-1em}

\subsection{Extensions} \label{sec:dv-extension}
The DV protocol is highly extensible, and we present several of them. To make them easier to understand, we try to present the extensions using Equation (\ref{eqn:packet_dv}) whenever possible.

It is important to note that despite the extensions below, there can be global requirements that are too complex or may not be able to be converted to \dnet; for example, multi-path consistency and disjoint-path~\cite{kazemian2012header, khurshid2013veriflow, kazemian2013real, yang2015real, horn2017delta} are such requirements. These requirements will be delegated to our FIB state distribution (FSD) protocol; see Appendix~\ref{sec:fsd}.  

\para{Handling packet modification.} An FIB entry can specify a modification function $f$. It takes as input $p$ and outputs a new packet. All FIB entries at $x$ compose a modification function $f_x$. The Equation (\ref{eqn:packet_dv}) becomes:
\begin{equation} \small
    d_p(x) = d_{f_x(p)}(n_p(x)).
\end{equation}

\para{Handling multicast and anycast.} An FIB entry can specify a group $N_p(x)$ of next-hops, in the scenario of anycast (load-balancing) and multicast. For multicast, the $d_p(x)$ is $True$ only if all of the nexthops' $d$ functions are true. Therefore, it can be simply specified using the $\bigwedge$(AND) function. 
\begin{equation} \small
    d_p(x) = \bigwedge_{y \in N_p(x)} d_p(y).
\end{equation}

The handling of anycast depends on the requirement. If the requirement is that any one of the next hop is acceptable, it will be handled in the same way as multicast. Otherwise, if one next-hop following requirements is enough,  Equation (\ref{eqn:packet_dv}) is written as:
\begin{equation} \small
    d_p(x) = \bigvee_{y \in N_p(x)} d_p(y).
\end{equation}

\para{Link-state check.} An FIB entry taking action to a next-hop with no link available should be considered as a violation.
We denote the state of the link from $x$ to $n_p(x)$ as $l_x(n_p(x))$ ($True$ means available), then we can integrate this variable to Equation (\ref{eqn:packet_dv}): 
\begin{equation} \small
    d_p(x) = l_x(n_p(x)) \wedge d_p(n_p(x)).
\end{equation}
Note that if the link to $n_p(x)$ fails, the node cannot receive $d_p(n_p(x))$. But the $d_p(x)$ is computed to be  $False$ since \\ $l_x(n_p(x)) = False$. 

\para{Conditional requirement.} The DV protocol can extend with predicates on variables (like link-state) as conditional requirements. For example, at $x$, a backup path can be specified with a next-hop $b$, it will only be used when the primal path (with the next-hop $t$) is not available. The Equation (\ref{eqn:packet_dv}) can be extended with predicates:
\begin{equation} \small
\begin{split} 
    d_p(x) = d_p(n_p(x)) \wedge (l_x(t) \Rightarrow (t = n_p(x))) \\ \wedge (\lnot l_x(t) \Rightarrow (b = n_p(x))).
\end{split}
\end{equation}

\para{Coverage requirement.} 
The DV protocol can verify not only whether a given path satisfies the requirements, but also the coverage  requirements (completeness). Consider the requirements that ``Intent 3. All redundant shortest paths should be available", specified for Azure in~\cite{jayaraman2019validating} using anycast. Let the set of redundant nexthops be $C(x)$. We have:
\begin{equation} \small
    d_p(x) = (\bigwedge_{y \in N_p(x)} d_p(y)) \wedge (C(x) \subseteq N_p(x)).
\end{equation}
Apply the condition to the Azure example, where $C(x)$ is all outgoing edges of $x$, notated as $C(x) = E_{out}(x)$. Also, we have $d_p(dst) = True$ and $\forall y \notin E_{out}(x)$, $d_p(y = nil) = False$ by definition. Then, we have the following derivation:
\begin{equation} \small
\begin{split} 
    \bigwedge_{y \in N_p(x)} & d_p(y) = (E_{out}(x) \supseteq N_p(x)) \wedge (\bigwedge_{y \in E_{out}(x)} d_p(y))\\
    d_p(x) &= (\bigwedge_{y \in E_{out}(x)} d_p(y)) \wedge (N_p(x) = E_{out}(x)) \\
    d_p(src) &= (\bigwedge_{y \in E_{out}(src)} d_p(y)) \wedge (N_p(src) = E_{out}(src))\\
             &= (N_p(src) = E_{out}(src)) \wedge (\bigwedge_{y \in E_{out}(src)} (N_p(y) = E_{out}(y))\\
             &  \wedge (\bigwedge_{z \in E_{out}(y)} d_p(z)))\\
             &= (\bigwedge_{y \neq nil} (N_p(y) = E_{out}(y))) \wedge d_p(dst)\\
             &= (y \neq nil) \Rightarrow (N_p(y) = C(y)).
\end{split}
\end{equation}
The above expansion of equation is a traversal on the \dnet, which always ends at $dst$. As a result, every node $y$ can detect violation using purely local contracts $N_p(y) = C(y)$.


\para{Controlling placement of verification.} 
Although the default deployment model of the DV protocol is to deploy it for each source, following fate-sharing, the framework is flexible to allow control of deployment. Fundamentally, each point of the DV protocol can verify (computing $d()$) for only the paths starting from the point. For example, Instead of having individual server as the source, since all servers' prefixes within a rack are aggregated at ToR switches, a deployment may deploy only at ToR switches, verifying the requirements from those points on.

\section{Distributed Packet Forwarding}\label{sec:composition}

After \DistVeri verifies the FIBs from each control plane efficiently, \textit{CP Composer} makes the CP selection and decides on how to route each individual packet using the results of \DistVeri. Assuming operators specify a rank among all CPs, \system has the ingress node select the highest-ranked verified CP, and announces it to other nodes. This is a simple yet powerful composition, as there is a single decision maker and it guarantees eventual consistency. 

However, there are two challenges. First, consider a single CP. When the source has verified that the CP has no violation and hence chosen it for a packet, the CP of a switch may have changed its FIB when the packet arrives at it. This changed FIB may be invalid. We refer to this as the \system\ \textit{consistency problem}. Secondly, to guarantee correctness, operators may require that only verified states of the data plane can be used. When intermediate nodes modify their FIBs, they have to wait for the verification results before being able to forward the packets, potentially leading to increased delays. We refer to this as the \textit{updating blocking problem}. Third, a CP may oscillate between various routes, causing \textit{CP Composer} to continuously recompute and update the CP selection for the same updates. We refer to this as the \textit{Unstable CP composition problem}.



\vspace{-1em}
\subsection{\system Forwarding Consistency}\label{sec:consistency}
\system allows users to specify and satisfy different consistency requirements (\eg, per-packet, eventual, loop-free) per packet space. However, there exists a trade-off between the level of complexity of the solution, and consistency requirements. By default, \system guarantees eventual consistency by having the ingress node select the highest-ranked verified CP, and announces it to other nodes. 
When traffic in the same packet space enters the network from multiple nodes, these ingress nodes run a consensus protocol 
to let one ingress device decide the CP to use.
\system can achieve stronger consistency requirements. In particular, the verifier (source in typical cases) 
can orchestrate the commitments of FIB changes using consistent updates, solutions which have also been developed to guarantee different types of consistency while minimizing the disruption time (\eg,~\cite{liu2013zupdate, jin2014dynamic}).
\newline

\vspace{-2em}
\subsection{Update Blocking Problem}

Link and node failures can incur delays to \system, as the nodes must complete the verification procedure before packets can be forwarded along the newly computed paths. This section introduces mechanisms to reduce disruption times.

Before presenting the mechanisms, we first describe the issue in further detail. To illustrate the problem, we consider a source $S$ sending traffic using a CP along the path $A$-$B$-$C$-$D$. The failure of the link $B$-$C$ may cause $B$ to select another node as its next-hop, \eg, $X$. However, before committing to the change, $B$ has to notify $S$, $S$ must complete the verification, and $B$ has to receive the notification from $S$ even if the updated path is correct. This blocking update might cause performance degradation due to the incurred delay. 

In order to mitigate the problem, \system introduces the fast reroute (FRR) ability to each control plane. Specifically, the backup paths are treated as virtual CPs, and pre-verified. As such, immediately upon detecting a network event, an intermediate node (\eg, $B$) may switch to the current CP's FRR. Backup paths have traditionally been implemented in two ways: alternate next-hops and tunneling. 

For alternate next-hops, a number of control planes support it for local failure protection on the data plane. For example, EIGRP and RIP provide loop-free alternates in the routing protocol~\cite{eigrp-rfc, rip-rfc}, and BGP maintains a list of policy-compliant routes in the RIBs where the second-highest rank one can be naturally considered as a backup route after being verified~\cite{BGP}. The backup next-hop can be verified by having it injected into a virtual CP, and executing \DistVeri before the targeted failures. 
Another common way of implementing FRRs for link protection is through tunnels~\cite{zheng2016we,liu2014traffic,wang2010r3}, and requirements for the tunneled path (\eg, waypoint avoidance) are also verified through a virtual CP.

In addition to these two common forms of implementation, 
\system{} takes advantage of the co-existence of multiple control planes and allows one CP to be used as the FRR mechanism for another CP. More specifically, let us assume a device with two control planes: $CP_1$, and $CP_2$ with $CP_1$ being preferred CP.
When the next-hop in $CP_2$ is different from that of $CP_1$, the next-hop is treated as and injected into a virtual CP of $CP_1$. After successful verification, the next-hop can then be used as a backup path for $CP_1$.

\vspace{-1em}
\subsection{Unstable CP Composition Problem}
In some instances, a CP may oscillate between different routes due to hardware, software or configuration errors, leading to continuous repeated updates to the vFIB. Because \DistVeri eagerly recomputes correctness results in response to any change in a control plane's vFIB, this may lead to a packet's CP assignment oscillate excessively as well.

Inspired by route flap damping in BGP, we can deploy a CP oscillation damping mechanism to address this issue. Each packet space is assigned a penalty value which is increased by a fixed parameter (\eg, 1000) every time the CP assignment is updated. If the penalty value exceeds a \textit{suppress limit} (\eg, 3000), any subsequent \DistVeri updates will not trigger the CP composer. The penalty value also decays exponentially according to a \textit{half-life time} (\eg, 5 minutes) as long as no \DistVeri updates are triggered. If the penalty for a suppressed packet space falls below a \textit{reuse limit}, the composer will resume responding to subsequent verification updates. The suppress limit, half-life time, and reuse limit may all be configurable.
This mechanism reduces the impact of oscillating vFIB updates on the CP assignment without affecting the reaction time and availability of stable routes.


\section{Evaluation}\label{sec:evaluation}
A switch OS level verification software suite called \impl is implemented to deploy \system{} on real white-box switches (See Appendix~\ref{sec:implementation} for details). 
This section illustrates the benefits of \system, evaluates its overhead under different network settings, and demonstrates its scalability and suitability even for large-scale networks. To this end, \system is evaluated extensively using 
both emulation and simulation on a variety of networks described in Table~\ref{tab:topologies}. Details of the experiment environments are described in Appendix~\ref{sec:eval-set}.

\begin{table}[t]
\setlength{\abovecaptionskip}{0.1cm}
\setlength{\belowcaptionskip}{-0.cm}
\centering
\footnotesize
    \begin{tabular}{|c|c|c|c|}
    \hline
    Network & Type & \# nodes & \# links \\
    \hline
    Stanford~\cite{stanfordnetwork}& Backbone & 16 & 14\\
    \hline
    AT\&T~\cite{attnetwork} & Backbone & 25 & 56\\
    \hline 
    Rocketfuel (AS 1755)~\cite{rocketfuel} & Backbone & 172 & 381 \\
    \hline 
    3-layer $k$-ary fat tree & Data center & $5k^2/4$ & $k^3 / 2$ \\ 
    \hline
    \end{tabular}
    \caption{\small{Summary of network topologies we used.}}
    \label{tab:topologies}
    \reducespace
    \vspace{-2em}
\end{table}

\begin{figure*}[t]
\setlength{\abovecaptionskip}{0.1cm}
\setlength{\belowcaptionskip}{-0.cm}
  \begin{tabular}{@{}c@{}c@{\hspace{1em}}c@{}}
     \includegraphics[width=0.7\columnwidth]{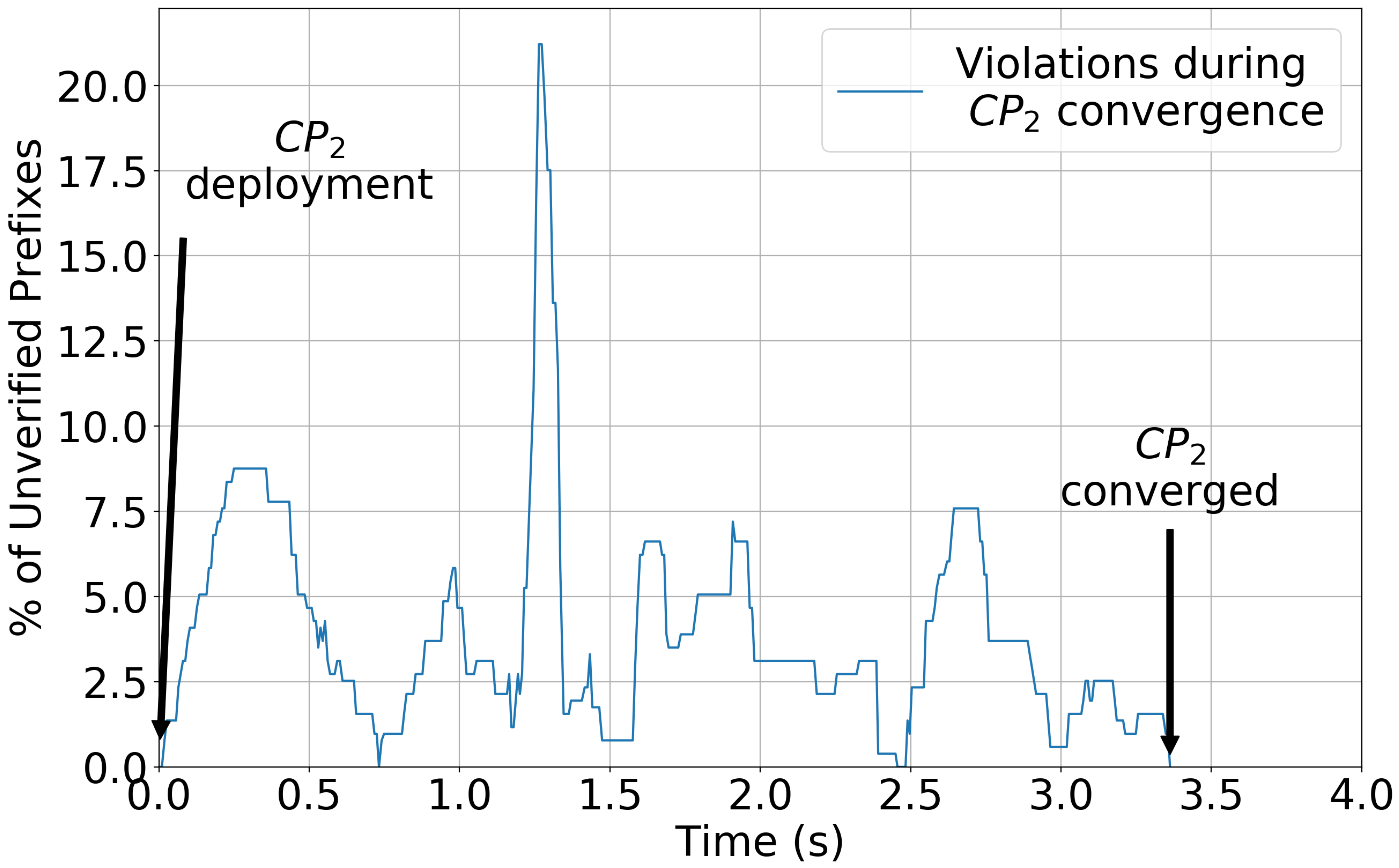} & \includegraphics[width=0.7\columnwidth]{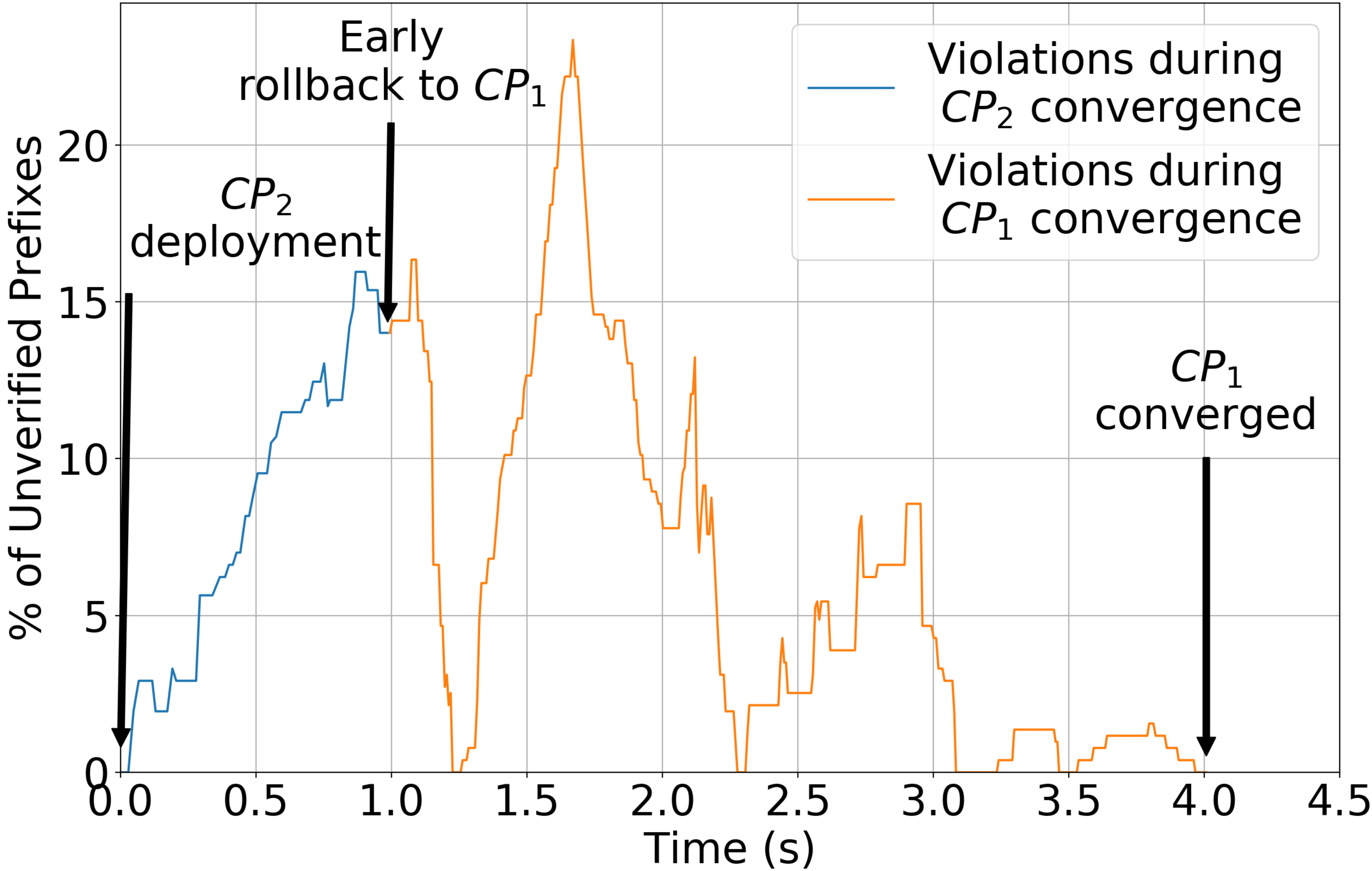} & \includegraphics[width=0.7\columnwidth]{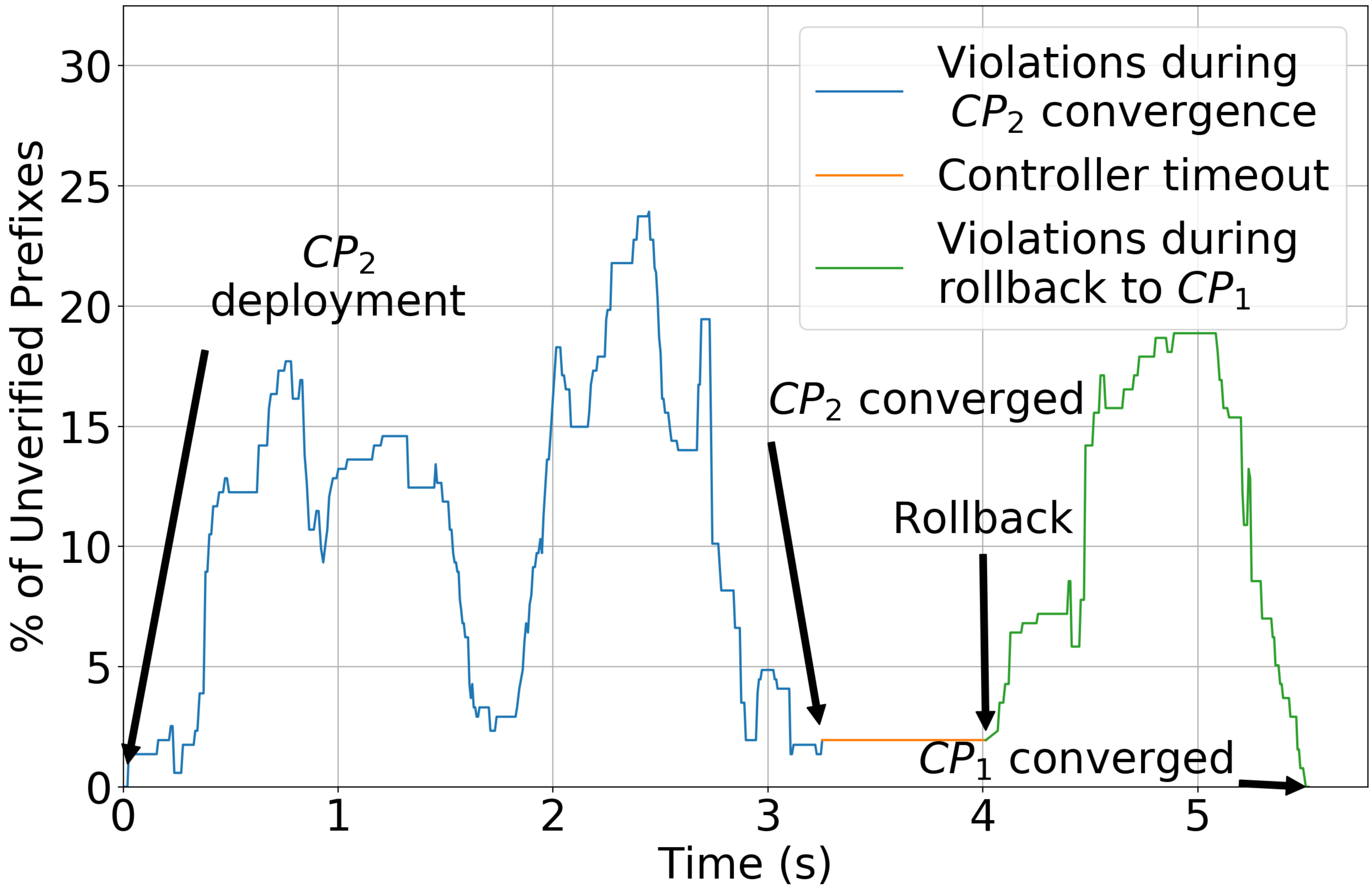} 
    \\ \small(a) & \small (b) & \small (c)
  \end{tabular}
    \caption{\small{The percentage of packet space that violates the correctness requirement when migrating from $CP_1$ to $CP_2$. (a) The newly deployed $CP_2$ does not have any bugs, yet, the network encounters transient violations during the migration. (b) Centralized verification (\ie, NetPlumber) can incorrectly detect data plane errors during CP convergence, resulting in a premature CP rollback. (c) $CP_2$ has bugs, and the network performs a global rollback to $CP_1$, resulting in significant interruption.
    }}\label{fig:coldstart}
    \vspace{-1em}
\end{figure*}


\subsection{Benefits of \system}
The goal of this set of experiments is to demonstrate the benefits of \system in ensuring correctness requirements under either software bugs or network failures.

\subsubsection{IGP Migration}~\label{sssec:coldstart} 
This first experiment demonstrates how \system can ensure that no policy requirements are violated even during operations that are highly susceptible to major disruptions. More specifically, we consider an IGP migration, \ie, the replacement of a network’s existing, or legacy, IGP with a different version of IGP. Upgrading the IGP can result in performance improvements and new service capabilities. However, the process can also lead to periods of severe network disruption (\eg, loops, congestion, blackholes). To mitigate the problems, several approaches have been developed (\eg, \cite{vanbever2011seamless,vanbever2012lossless}).

We show how \system can ensure that the policy requirements are still satisfied during IGP migrations. In particular, we assume that the administrator wants not only loop-free forwarding paths, but also packets to be forwarded along the shortest paths 
to reduce the risks of congestion. We compare the performance of \system with the overlay method~\cite{vanbever2012methods,fboss}, a best current practice approach for IGP migration, enhanced with a central verification server. In the overlay method, all routers run not only the legacy IGP, but also the new IGP concurrently, with routers initially preferring the legacy IGP. Incrementally, the configurations of the routers are updated for the new IGP to be preferred. During this period of transition where a number of routers prefer the legacy IGP, and others prefer the new IGP, undesirable routing outcomes may happen: packets may result in forwarding loops, blackholes, or get forwarded along long convoluted paths. 

We run the experiments on a network topology emulating that of the AT\&T backbone shown in Table~\ref{tab:topologies}.
To measure the number of policy violations, we deploy a server running NetPlumber~\cite{kazemian2013real}. Whenever a FIB rule is added, it is immediately sent to the real-time centralized verification controller which can then identify and count the number of violations of basic reachability, loop-freeness, and shortest path.



\para{Results.} Figure~\ref{fig:coldstart} (a) shows the percentage of packet space that violates reachability, loop-freeness or shortest path during the upgrading process, computed by NetPlumber. We observe that up to 21\% of the packet space may result in violations in the network. In contrast, with \system, the percentage is 0\% throughout the process. 

In response to detecting violations, a number of proposals (\eg,~\cite{tilmans2014igp, vissicchio2015co, ebb, jain2013b4,fboss}) have advocated rolling back to the legacy IGP version. In particular, the new version of the software may contain errors (\eg, bugs). However, because centralized verification can only verify a snapshot of a data plane, the snapshot may represent a transient state, and the server may unnecessarily rollback the software causing further network disruptions. Figure~\ref{fig:coldstart} (b) illustrates the problem: After deploying the new IGP, transient violations during the upgrade trigger a software rollback to the previous version of the IGP, resulting in a more severe violation. 
  
As another variant of the setting, we consider the new IGP software indeed has a bug. As shown in Figure~\ref{fig:coldstart} (c), there are still 2\% of the packet space that experience violations even after the migration has been completed. Because of the bug, the IGP at every router in the network has to be roll backed to the previous version. The results demonstrate significant disruption and violations during the process. 
In contrast, \system always guarantees 0-violation even with a faulty new IGP since \system only allows the verified rules coming from the new IGP to be deployed in the data plane. This set of experiments demonstrates the difficulties in migrating IGP and shows the benefits of \system in such process.

\begin{table}[t]
\setlength{\abovecaptionskip}{0.1cm}
\setlength{\belowcaptionskip}{-0.cm}
\centering
\footnotesize
    \begin{tabular}{|c|c|c|c|}
    \hline
    ~ & SDN & \system & OSPF \\
    \hline
    Before partition & 171/171 (100\%) & 171/171 (100\%) & 105/171 (61\%)\\
    \hline
    After partition & 72/122 (59\%) & 115/122 (94\%)& 69/122 (56\%)\\
    \hline
    \end{tabular}
    \caption{\small{Fraction of flows that can traverse  the waypoint.}}
    \label{tab:table-partition}
    \reducespace
    \vspace{-2em}
\end{table}

\subsubsection{Waypoint Routing} 
This experiment continues to evaluate the effectiveness of \system in ensuring correctness requirements, but we now consider other objectives. More specifically, we consider the requirement of waypoint routing. Administrators commonly want specific traffic to traverse different network functions. 

We use the Rocketfuel topology in Table~\ref{tab:table-downtime}, and select one node as the destination, another node as the waypoint to be traversed, and consider the paths from all other nodes. We compare three settings: First, switches run SDN only. Second, switches run OSPF only. Finally, switches run SDN as the preferred and primary CP, and OSPF as a backup CP. We gradually fail random links, until a subset of nodes gets partitioned and loses connectivity with the SDN controller. For each setting, we report the fraction of nodes that are in the partition with the target destination and waypoint node and that can satisfy the objective.


\para{Results.} Table~\ref{tab:table-partition} shows the fraction of paths that satisfy the requirement. Before the network is partitioned, it has of 171 source nodes. 
Using SDN, all of the paths satisfy the requirement. Similarly, with \system, the requirement is satisfied for all paths. In contrast, with OSPF, only 61\% of paths traverse the waypoint. After network partition, the component disconnected from the SDN controller has of 122 nodes. 
 
Using SDN, only 59\% of the paths satisfy the requirement. This is because the OpenFlow rules may be obsolete, and may forward packets along invalid paths and into black holes. 
Using OSPF, the fraction of paths that satisfy the requirement drops to 56\%, and the remaining 44\% would still allow packets to reach the destination but without traversing the waypoint, and would thus violate correctness. 
In contrast, with \system, 94\% of the paths satisfy the requirement, and for the remaining 6\%, their traffic would be dropped. As such, \system guarantees correctness for all of the paths.

\begin{figure}[t]
\setlength{\abovecaptionskip}{0.1cm}
\setlength{\belowcaptionskip}{-0.cm}
\begin{tabular}{@{}c@{}c@{\hspace{1em}}c@{}c@{}}
\centering
\includegraphics[width=0.5\linewidth]{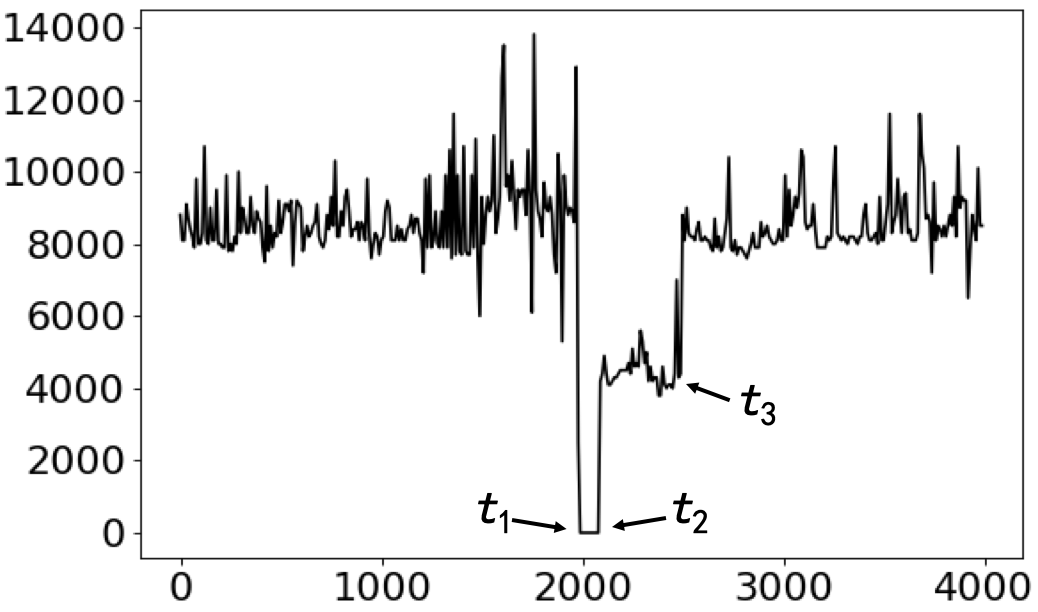} &
\includegraphics[width=0.5\linewidth]{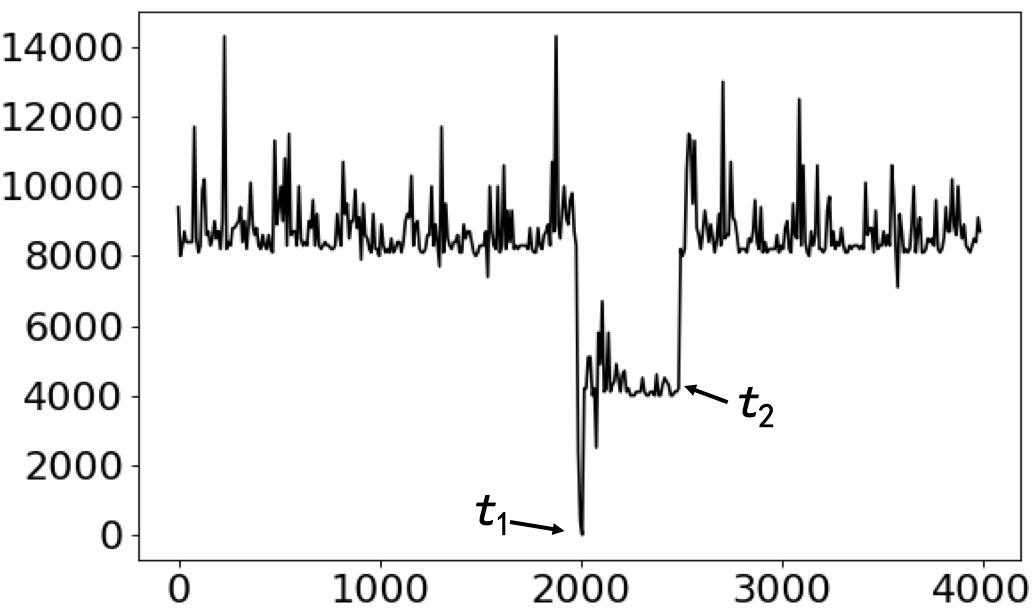} \\
\small(a) & \small(b)
\end{tabular}
\caption{\small{Packet receiving rate for the fast recovery experiments. The failure in (a) affects both the SDN and OSPF CPs, while that in (b) affects only the current SDN. }}\label{fig:sdn-ospf1-sdn}
\reducespace
\vspace{-1em}
\setlength{\abovecaptionskip}{-1em}
\end{figure}

\subsubsection{Failure Recovery} This experiment demonstrates the effectiveness of \system in recovering from link failures.
This experiment is under the eventual consistency model as defined in Section~\ref{sec:consistency}, and enables FRR across CPs, \ie, upon detecting a link failure, a node can decide to switch to another CP which is not using that link. We deploy a Rocketfuel topology in Table~\ref{tab:table-downtime}, and two of the routers are replaced with the \emph{white-box switches}. Each of the rest network devices runs as a separate Docker container, with two CPs: a Quagga OSPF, and an SDN.

We randomly select a pair of nodes and generate UDP traffic between them by running iperf~\cite{iperf} at a rate of 100 Mbps.
We then randomly select links to fail on one of the paths used by the UDP traffic.
For the SDN CPs to recover from the failures, we implement a reactive approach. That is, when a failure happens, the device that detects it sends a control message to the controller to recalculate alternative forwarding paths and update the vFIB of the affected devices accordingly~\cite{sharma2012demonstration}. 
For each run, we measure the downtime defined as the amount of time between the moment the destination stops receiving packets because of the failure to the moment the receiver starts receiving packets again. We repeat this process for 10 runs.

\para{Results.} 
Figure~\ref{fig:sdn-ospf1-sdn} illustrates the patterns we observed: First, in Figure~\ref{fig:sdn-ospf1-sdn}(a), the failure affected both SDN and OSPF CPs. In other words, both CPs used the failed link for the next hop. As such, at $t_1$, no valid route can be used, leading to a throughput of 0. Then, after OSPF discovers an alternate path, \system uses OSPF to forward traffic, and the throughput starts increasing at $t_2$. Finally, after the SDN recovers the route, the device uses the OpenFlow rules to forward traffic, and the rate increases further starting from $t_3$. In Figure~\ref{fig:sdn-ospf1-sdn}(b), the link failure affects only the SDN CP. The device can immediately switch to OSPF using cross-CP FRR to continue forward the packets. 

Table~\ref{tab:table-downtime} summarizes the average downtimes of the two CPs, and that of \system{} across the runs.
While the average downtime of the SDN CP is 112.861 ms, the average downtime of OSPF is 22.572 ms, and that of \system is even lower than both of them with 8.492 ms.
The results show that \system can reduce the downtime of SDN by more than an order of magnitude, and that of OSPF by 43\%. 
While one may be surprised that on average, \system{} can recover faster than OSPF, the main reason is that in some runs, the SDN CP recovers faster, whereas, in other runs, the OSPF CP recovers faster; and in every case, \system running the two CPs switches to the one that recovers the fastest.

\begin{table}[t]
\setlength{\abovecaptionskip}{0.1cm}
\setlength{\belowcaptionskip}{-0.cm}
\centering
\footnotesize
    \begin{tabular}{|c|c|c|c|}
    \hline
    ~ & SDN & \system & OSPF \\
    \hline
    Average downtime & 112.861 ms & 8.492 ms & 22.572 ms \\
    \hline
    \end{tabular}
    \caption{\small{Average downtime of using different systems.}}
    \label{tab:table-downtime}
    \reducespace
    \vspace{-1em}
\end{table}

\vspace{-1em}
\subsection{Overhead and Scaling}
A concern of \system is its overhead 
determining the cost and the scaling ability of \system. Compared with a traditional network with a single CP and no distributed verification, \system has the following overhead: (1) the overhead of running each additional CP instance at each switch; and (2) the overhead in \system (especially from distributed verification) in terms of memory, messaging and processing. In this section, we first evaluate the feasible number of CPs we can run on a real network. We then extensively evaluate the overhead introduced by the \system using real backbone and data center topologies. \\


\vspace{-1.5em}
\subsubsection{Control Plane Overhead} 
This set of experiments aims to demonstrate \system's feasibility by a stress test that runs 1 to 150 BGP instances in each switch. For each experiment, we measure the network convergence time as the time the last BGP instance converges. Specifically, we emulate the Stanford backbone network~\cite{stanfordnetwork} on the VM having 32 vCPUs, and we limit each container to 2 vCPUs and 4 GB memory.~\footnote{Most of the modern switches have at least 2 GB of RAM and CPUs with two cores.~\cite{ciscohw1,juniperhw1}.} To have a similar number of FIB entries in the Stanford network, we connect each node to 3K external routes, making an average of 48K FIB entries in each node and ~750K prefixes in total. We then run different numbers of Quagga's BGP instances to understand how many CPs \system can support. We instantiate each BGP instance on a node with a different port and a private AS number. 

\begin{figure}[t]
\setlength{\abovecaptionskip}{0.1cm}
\setlength{\belowcaptionskip}{-0.cm}
\centering
\includegraphics[width=0.7\linewidth]{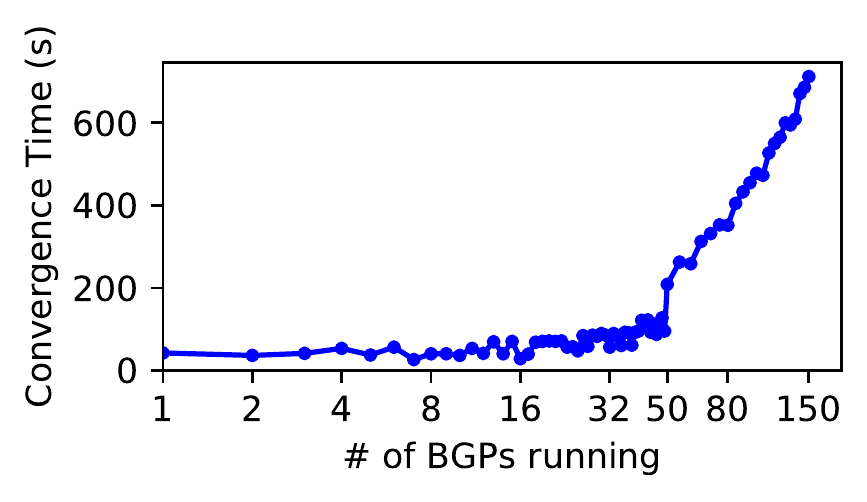} 
\caption{\small{Network convergence times of \system with different numbers of BGP instances per device.}}\label{fig:overhead}
\reducespace
\vspace{-1.5em}
\end{figure}

\para{Results.} Figure~\ref{fig:overhead} shows the network convergence time with different numbers of BGP instances running in the network. We observe that running multiple CPs on a commercial switch incurs a moderate overhead. Specifically, when the total number of BGP instances is less than 50, the network does not experience a significant increase on convergence time~(\ie, 50 seconds running one BGP vs. 100 seconds running 50 BGPs). However, when the total number of BGP instances exceeds 50, the convergence time increases non-linearly. We analyze the memory, bandwidth and CPU usages for each experiment, and conclude that the main bottleneck for running more than 50 BGPs in the experiment is insufficient CPU resource, causing high scheduling delays. 



\vspace{-2mm}
\subsubsection{Verification Overhead} 
In this set of experiments, we evaluate the DV protocol on various backbone and data center networks. By gradually scaling topology size and varying the number of FIB entries, we demonstrate how well \system scales. For backbone networks, we select topologies of different sizes from~\cite{knight2011internet} to run OSPF, and control the number of external routes announced by each router. After OSPF converges, we randomly pick an FIB entry to modify and then measure the messaging and processing overhead. For data center (DC) networks, we use 3-layer k-ary fat tree models with different $k$ values (from 24 to 80). We simulate the protocol execution on involved nodes and estimate the overall overhead. As for requirements, we use reachability and loop-freeness for backbones, and shortest path and reachability for data centers. 

\begin{figure}[t]
\setlength{\abovecaptionskip}{0.1cm}
\setlength{\belowcaptionskip}{-0.cm}
\begin{tabular}{@{}c@{}c@{\hspace{1em}}c@{}c@{}}
\centering
\includegraphics[width=0.5\linewidth]{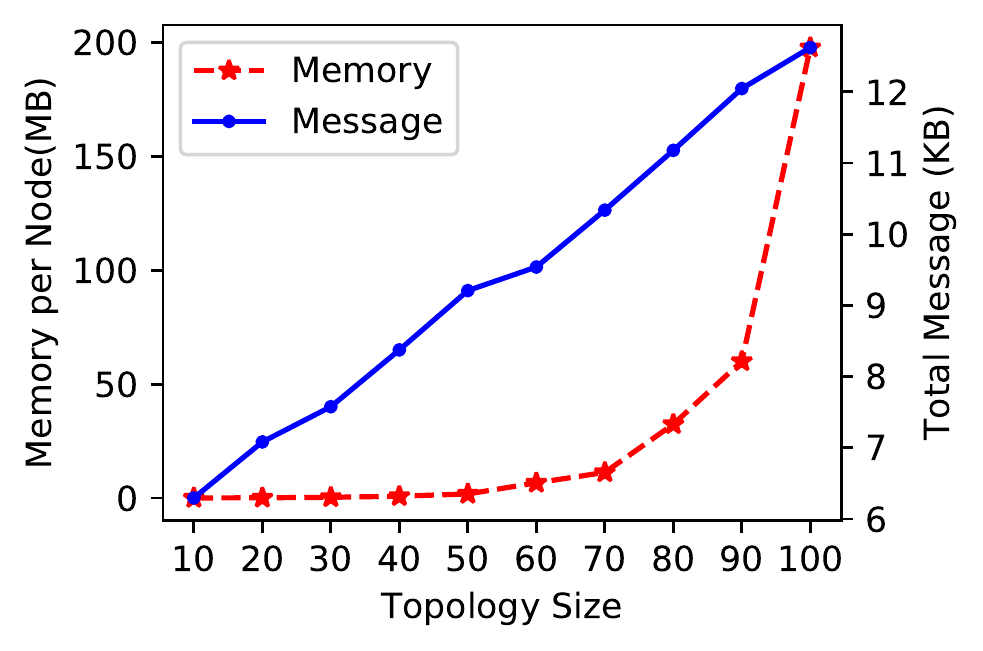} &
\includegraphics[width=0.5\linewidth]{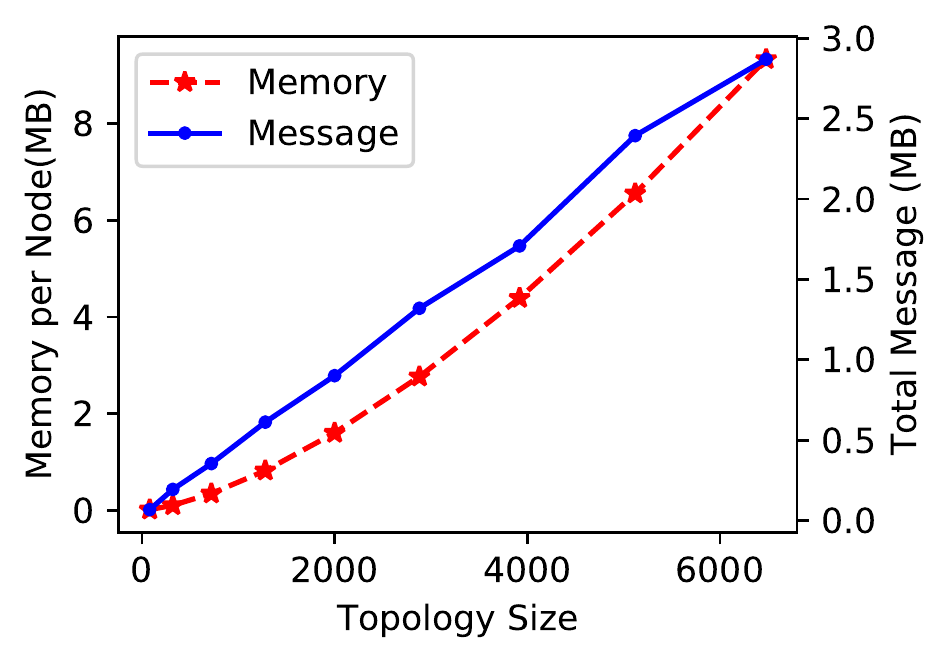} \\
\small(a) & \small(b)
\end{tabular}
\caption{\small{Average memory consumption per device and total message after a FIB update. (a) Backbone (b) Data center.}}\label{fig:single_update}
\reducespace
\end{figure}

\para{Memory overhead.}
Figure~\ref{fig:single_update} shows how memory usage of each node grows with topology size. We observe that the per-device memory consumption of the DV-protocol in backbone networks reaches 190 MB when the topology size becomes 100. In contrast, 
the DV protocol only consumes 13 MB of memory on a DC network with 8K nodes. This is because the shortest path requirement substantially reduces the size of the \dnet.
With this requirement, \dnet for a single source contains the same number of nodes as the physical topology. Further, each device only needs to maintain the information about its upstream and downstream devices, requiring $O(n^{3/2})$ memory for a topology of size $n$.
Considering that modern switches normally have above 1 GB memory, we draw the conclusion that the memory overhead of the DV-protocol is is insignificant in DC networks, and is moderate in backbone networks.

\para{Messaging overhead}. 
Figure~\ref{fig:single_update} illustrates the amount of message traffic incurred by the DV-protocol in the whole network after one device updates its verification result.
We observe that the DV-protocol incurs a negligible message overhead, which does not limit its scalability: the total traffic does not exceed a few MBs even when the network has 8K network devices. Both backbone and data center topologies, moreover, have a linear growth of the message traffic with the topology size. This is because the filtering nature of \dnet design substantially reduces redundant messages while the message overhead of the naïve FIB flooding is proportional to the number of links in the network.

\begin{figure}[t]
\setlength{\abovecaptionskip}{0.1cm}
\setlength{\belowcaptionskip}{-0.cm}
\begin{tabular}{@{}c@{}c@{\hspace{1em}}c@{}c@{}}
\centering
\includegraphics[width=0.48\linewidth]{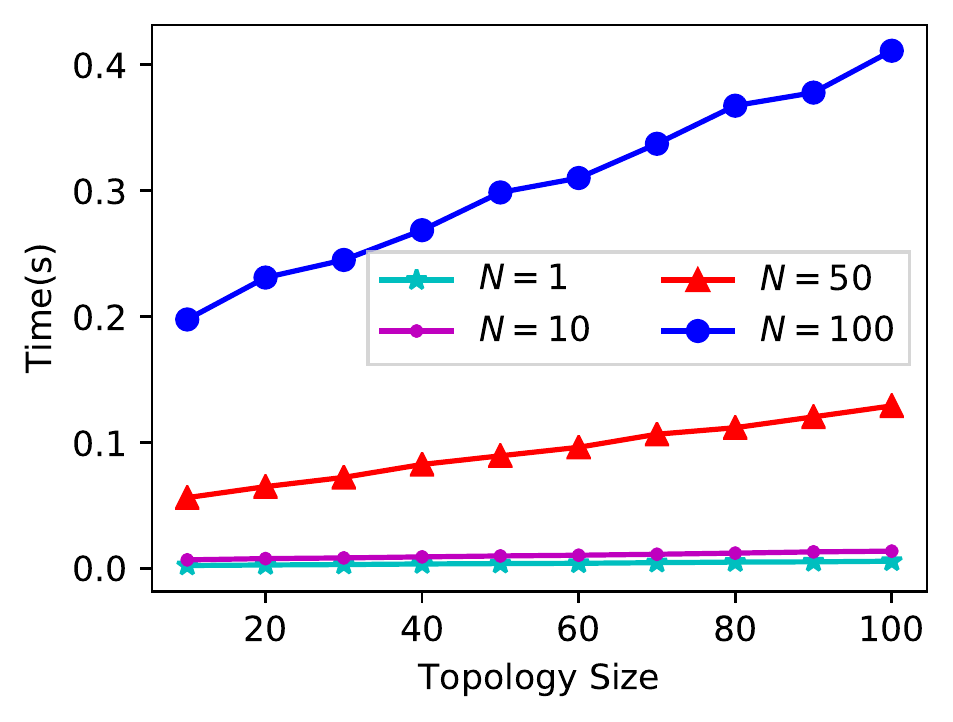} &
\includegraphics[width=0.50\linewidth]{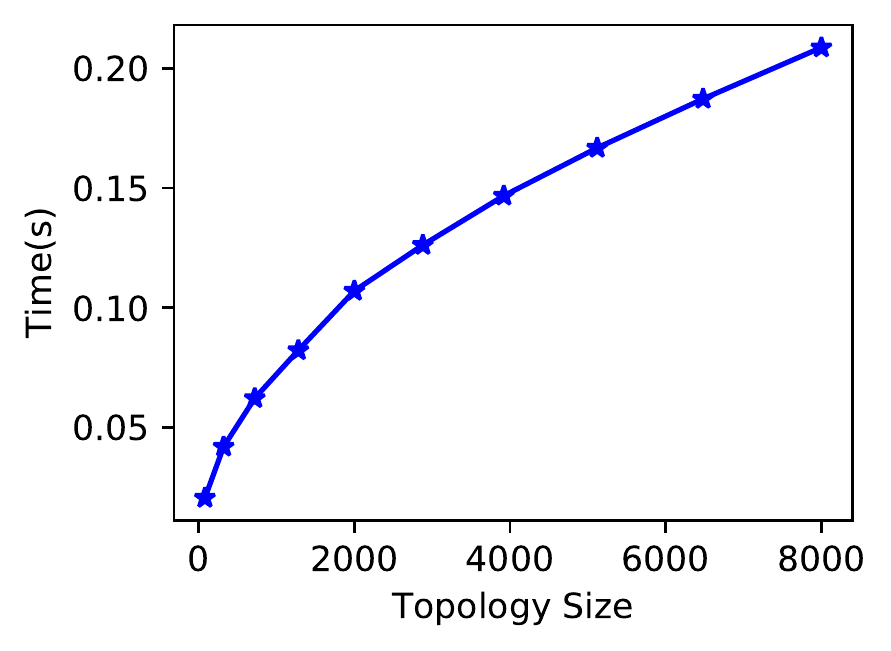} \\
\small(a) & \small(b)
\end{tabular}
\caption{\small{Processing time of \system for a single FIB update. (a) Backbone (b) Data center. $N$ is the average number of external routes per device.}}\label{fig:processing}
\reducespace
\vspace{-1em}
\end{figure}

\para{Processing overhead}. 
Figure~\ref{fig:processing} shows the total time required for a device's verification function update to be propagated to and get processed at the corresponding source.

Figure~\ref{fig:processing}(a) shows that the processing time increases as the topology size and the number of external routes announced by each router on the backbone networks increase. The processing time reaches 0.4 s on a topology with 100 nodes where each node announces 100 external routes. Essentially, the hop count and the average number of FIB entries determine the total processing time of the DV-protocol.

Figure~\ref{fig:processing}(b) plots the processing time of the DV-protocol in DC networks collected using numeral simulation. In the simulations, the number of external routes increases linearly as $k$. We observe that although the processing time increases with the size of the DC network, the increasing speed becomes slower as the network size keeps increasing. As such, the DV protocol has a small processing overhead.




To summarize, \system, with its light-weight design and message pruning mechanism, can scale effectively to perform distributed verification for large networks.

\vspace{-2mm}
\section{Related Work}\label{sec:related-work}

\para{Network verification}. 
A number of methods and tools have been proposed to verify network behaviors
\cite{kazemian2012header, khurshid2013veriflow, kazemian2013real, yang2015real, horn2017delta, beckett2017general, gember2017automatically,
handigol2014know, zeng2012automatic, batfish}. These tools adopt a common design, which is to use a 
centralized server to collect network forwarding information or configurations from all network devices, and analyze network forwarding behaviors. This design requires reliable connections between the server and network devices, and makes the server a performance bottleneck and a single point of failure. Azure~\cite{jayaraman2019validating} allows devices to locally verify its forwarding behavior using local contracts, but is limited to only verify shortest path reachability and fault-tolerance. In contrast, \system provides \DistVeri{}, a simple, generic, efficient distributed verification framework that can verify a wide range of common requirements (\eg, reachability, waypoint, subnet isolation, loop-freeness, and fault-tolerance).



\para{Network configuration synthesis and emulation}. 
Network configuration synthesis~\cite{merlin, contra, propane, el2017network, el2018netcomplete} and emulation~\cite{liu2017crystalnet,lopes2019fast, fboss} are complementary to verification. Network synthesis~\cite{merlin, contra, propane, el2017network, el2018netcomplete} lets operators specify high-level intent, and generates corresponding configurations (\eg, routing protocol configurations and programmable data plane). Network emulation~\cite{liu2017crystalnet,lopes2019fast, fboss} emulates the execution of these configurations before they are deployed in the networks. They aim to build a single infallible control plane, which often results in an over-engineered, complex control plane or a control plane with limited capability. Rather than relying on a single control plane, \system{} resorts to an "Internet way" to run multiple control planes and hot-swap among them to improve network reliability. In \DistVeri{}, one approach to transform a generic network requirement verification to reachability verification on \dnet{} is to use the product graph~\cite{mendelzon1995finding}, which is also used in Merlin~\cite{merlin}, Propane~\cite{propane} and Contra~\cite{contra} to synthesis configurations. 

\para{Multiple control planes}. 
Several designs have been proposed for composing multiple control plane layers to improve network reliability~\cite{kvalbein2006fast, tilmans2014igp, vissicchio2015co, ebb, jain2013b4, fboss, le2010theory}. For example, B4~\cite{jain2013b4} builds layered control architecture to run central TE on top of the baseline routing protocols, and FBOSS~\cite{fboss, ebb} agents run specific control planes (\ie, OpenR~\cite{openr} and BGP) simultaneously to compose both protocols' features. However, existing multiple control planes composition designs can only guarantee a limited number of properties (\eg, reachability and domain backup), but fail to provide generic routing correctness guarantees (\eg, waypoint routing, loop-freeness and subnet isolation). In contrast, \system{} uses the verification results of \DistVeri{} to systematically compose multiple control planes to guarantee generic routing correctness and operator-specified consistency models.

\vspace{-2mm}
\section{Conclusions}\label{sec:conclusions}
This paper presents \system{}, a novel system to
achieve high reliability in networks through distributed verification and multiple control plane composition. \system develops a simple, generic, distributed verification framework, and achieves systematic composition of multiple control planes using the verification results. Extensive experiments are conducted to demonstrate the benefits, overhead, and scalability of \system{}.

\newpage
\bibliographystyle{acm}
\bibliography{all}

\newpage
\clearpage
\pagenumbering{arabic}
\begin{appendices}
\section{A Generic FIB State Distribution Protocol}\label{sec:fsd}

This appendix presents \FSD, a novel FIB state distribution protocol that allows ingress devices to verify complex global requirements. 


The benefits of \FSD are twofold: (1) by allowing for \textit{purely distributed} verification, \FSD allows \system to enforce control plane requirements in spite of network partitions and failures, and (2) real-time verification allows \system to quickly assign control planes to incoming packets in response to network events. To achieve (1), we design lightweight data structures and novel algorithms to initialize and update them. To achieve (2), we utilize two key observations: (a) most data plane events only affect small subsets of the packet space, and (b) most data plane events only affect small subsets of network devices.

\FSD consists of two major components:
\begin{itemize}[leftmargin=*]
    \item A lightweight \textit{Local Equivalence Class Table} (LEC table) at each device to store relevant local forwarding information (\S\ref{dpv-sss1}).
    \item An efficient distributed update protocol to update the information at relevant devices in response to data plane updates (\eg, failures, configuration changes) (\S\ref{dpv-sss3}).
\end{itemize}
As each CP is associated with an independent \FSD instance, we focus on one single CP and its associated \FSD for the rest of this section. 

\subsection{LEC Tables}\label{dpv-sss1}
At each device, \FSD consists of local data structures known as Local Equivalence Class Tables (LEC Tables) which store the relevant forwarding behavior for outgoing flows at each device. To motivate this design, we first consider a naive distributed implementation of existing centralized verification systems (\eg, VeriFlow~\cite{khurshid2013veriflow}, NetPlumber~\cite{kazemian2013real}) by replicating the global data structures on each device. Devices would maintain the data structures by broadcasting any local flow rule updates across the network, and each device would individually compute correctness as specified by the centralized verification system. However, this approach has two major limitations: (1) the memory requirements of these global data structures generally scale with the \textit{total} forwarding rule space of the \textit{entire} network, which does not scale well in larger networks, and (2) network devices often have much less computing power than a typical controller, which may hurt verification time.

\begin{figure}[t]
    \setlength{\abovecaptionskip}{0.1cm}
    \setlength{\belowcaptionskip}{-0.cm}
    \centering
    \includegraphics[width=\columnwidth]{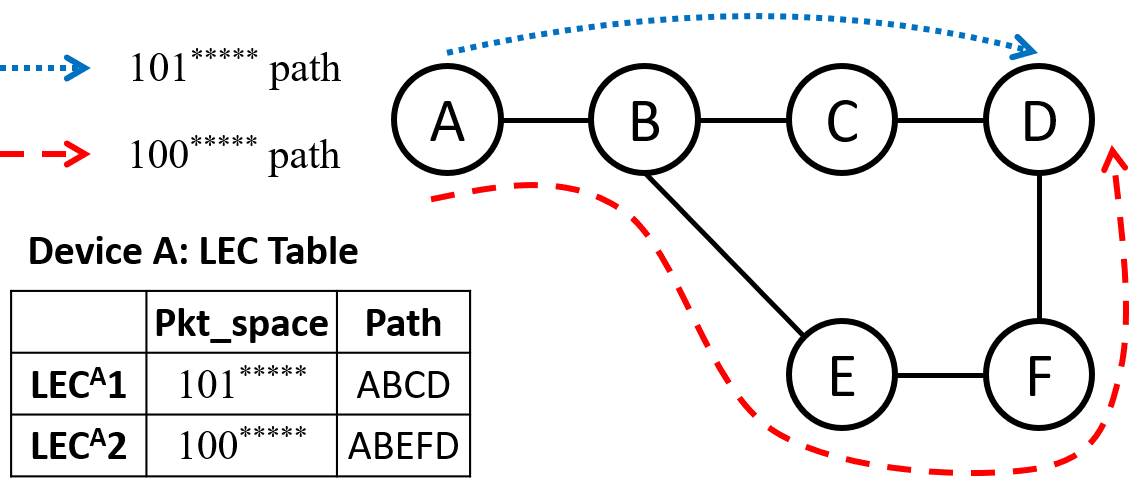}  
    \caption{\small{An example with two flows using $ABCD$ and $ABEFD$.}}\label{dpv-topo}

\end{figure}

To avoid these limitations, we observe that each device only needs to be aware of forwarding rules at subsequent devices that affect its outgoing flows. We can then trim the stored data by only considering the subset of network devices reachable by downstream flows. To illustrate these insights, consider Figure \ref{dpv-topo}. Any forwarding rules at device $E$ or device $F$ will not affect outgoing flows from device $C$, so device $C$ need not be aware of such forwarding rules. To take advantage of this, we will introduce the notion of Local Equivalence Classes. 


\begin{definition}[Local Path] The local path of packet $p$ at device $n$ ($lp_i^{n}(p)$) for a given control plane $CP_i$ is the downstream path of $p$ beginning at $n$ when $CP_i$ is stable.
\end{definition}



\begin{definition}[Local Equivalence Class] At each device $n$, the control plane $CP_i$ partitions the packet space into local equivalence classes $\mathcal{LEC}_i^n = \{\mbox{LEC}^n_j\}$, such that only packets whose headers are in the same LEC use the same \textit{unique} downstream forwarding path. Formally, for two arbitrary packets $p_1$ and $p_2$,
\end{definition}
\begin{itemize}[leftmargin=*, topsep=0.25em]
    \item \makebox[0.6cm][l]{$\forall j$,} \ $p_1,p_2 \in LEC^n_j$, then $lp_i^n(p_1)=lp_i^n(p_2)$; and
    \item \makebox[0.6cm][l]{$\forall j, k$,} \ $p_1 \in LEC^n_j, p_2 \in LEC^n_k$, then $lp_i^n(p_1) \neq lp_i^n(p_2)$.
\end{itemize}


At each device, \FSD partitions the packet packet space into LECs, and by definition, each LEC will have a unique downstream forwarding path. This associative map between LECs and forwarding paths will form LEC tables, the core data structures at each device, and this data will provide a complete local context on which \FSD can fully determine a given packet's forwarding behavior and thus allow us to query for general correctness requirements. Figure \ref{dpv-topo} shows the initial LEC table stored at device $A$.

To initialize the data structures, we utilize a vector-based algorithm, shown in Algorithm \ref{dpv-init}. Each device initially only has access to local forwarding information, which may only consist of next-hops (\eg, EIGRP). The initialization algorithm allows \FSD to aggregate next-hop information across devices in order to generate the LEC paths at each device.
\begin{algorithm}[t]
\small
\SetKwBlock{OnEvent}{on event}{}
\caption{LEC Table Initialization}\label{dpv-init}
\DontPrintSemicolon
\KwData{Local forwarding rules at each device}
Initialize associative map $T$\;
\tcc{First we initiate messages for LECs for which we are the end of the route}
\ForEach{flow rule f}{
    \If{f.nexthop == f.dst} {
        send LEC Path announcement ($n$, $f.match$, $n$) to all neighbors\;
        $T$.insert($f.match$, $n$)\;
    }
}
\OnEvent(incoming LEC path announcement $n_i$, $PS_i$, $P_i$){
    \ForEach{local flow rule f}{
        \If{f.nexthop == $n_i$}{
            $LEC^n_i \leftarrow f.match \cap PS_i$\;
            $T$.insert($LEC^n_i$, $P_i$)\;
            send LEC Path announcement ($n$, $PS_{end}$, $n \cup P_i$) to all neighbors\;
        }
    }
}
\end{algorithm}

\subsection{Distributed Real-time Updates}\label{dpv-sss3}
In this section, we will describe a novel distributed algorithm to verify correctness in response to real-time data plane updates. 
We will first introduce the three types of message utilized in \FSD, then demonstrate an example workflow in response to forwarding rules being added, modified, or deleted, and to other network events (\eg, link failures).

\para{LEC path update.} The goal of this message is to announce changes in a device's local forwarding state, as the changes may affect other devices' LEC paths. Two types of network events can trigger local forwarding state change: (1) forwarding rule update, and (2) topology change (\eg, link failure). 
When detecting such changes, \FSD first updates its local LEC table, and then generates an \textit{LEC path update message}, $(n, PS_{aff}, P)$, where $n$ is the device at which the network event was detected, $PS_{aff}$ is the packet subspace whose behavior is affected by the update, and $P$ is the new local path taken by $PS_{aff}$ starting at device $n$. If there is no longer a local path, $P$ is set as $NULL$. This announcement is broadcast to all devices in the network. Then, when receiving this announcement, each device performs a local search to identify affected LECs and updates its path information. To achieve this, the device checks for all LECs with a \textit{path dependency} on $n$, or in other words, all LECs whose forwarding paths contain $n$. In addition, the device checks for LECs with a \textit{packet space dependency} on $PS_{aff}$, or in other words, LECs that overlap with $PS_{aff}$. For each LEC that has both path and packet space dependencies on the incoming announcement, the device then updates its path according to the new segment $P$. The update may also partition an LEC into multiple ones, or even generate new LECs based on the intersection results.

\para{LEC request.} The goal of this message is for a device to retrieve LEC forwarding information from other devices. More precisely, in some instances such as when a forwarding rule is added or modified, a device may not have the needed local path $P$ stored in its LEC table. In such case, the device $n$ sends an \textit{LEC request message}, $(n', PS_{req})$, to the next-hop $n'$ of the new rule to request the relevant path of the match $PS_{req}$ of the new rule. 

\para{LEC reply.} In response to the LEC request message, device $n^\prime$ will search for the local path of $PS_{req}$ by finding the intersection in its own LEC table. Based on the LEC intersection results, $PS_{req}$ maybe split into multiple packet spaces $\{PS^i_{reply}\}$, and the union of all $PS^i_{reply}$ should equal to $PS_{req}$, \ie, $\ \bigcup\nolimits_{i} PS^i_{reply} = PS_{req}$. The LEC reply message follows the format as $(n', PS^i_{reply}, P^i)$ for each $PS^i_{reply}$, where $P^i$ is the local path of $PS^i_{reply}$ at device $n'$. Again, if there is no local path, $P^i=NULL$. After receiving the reply messages, $n$ will update its LEC table and broadcast the updated LEC messages.

We will now describe a realistic example to illustrate the behavior of \FSD in response to various network events as below. 

\begin{figure*}[ht]
\setlength{\abovecaptionskip}{0.1cm}
\setlength{\belowcaptionskip}{-0.cm}
  \centering
  \begin{tabular}{|@{\hskip 0.5em}c@{\hskip 0.5em}|@{\hskip 0.5em}c@{\hskip 0.5em}|}
    \hline
    \includegraphics[width=0.8\columnwidth]{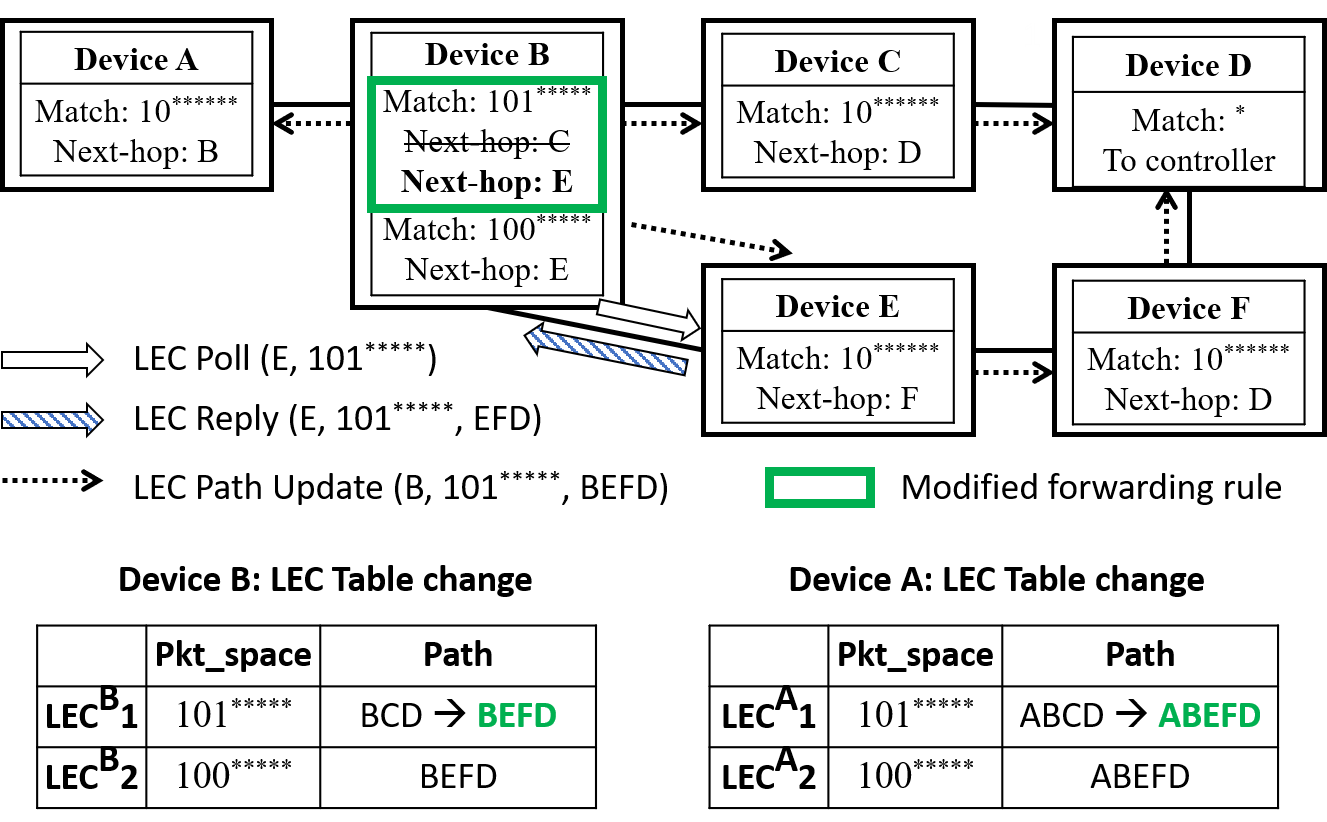} & \includegraphics[width=0.8\columnwidth]{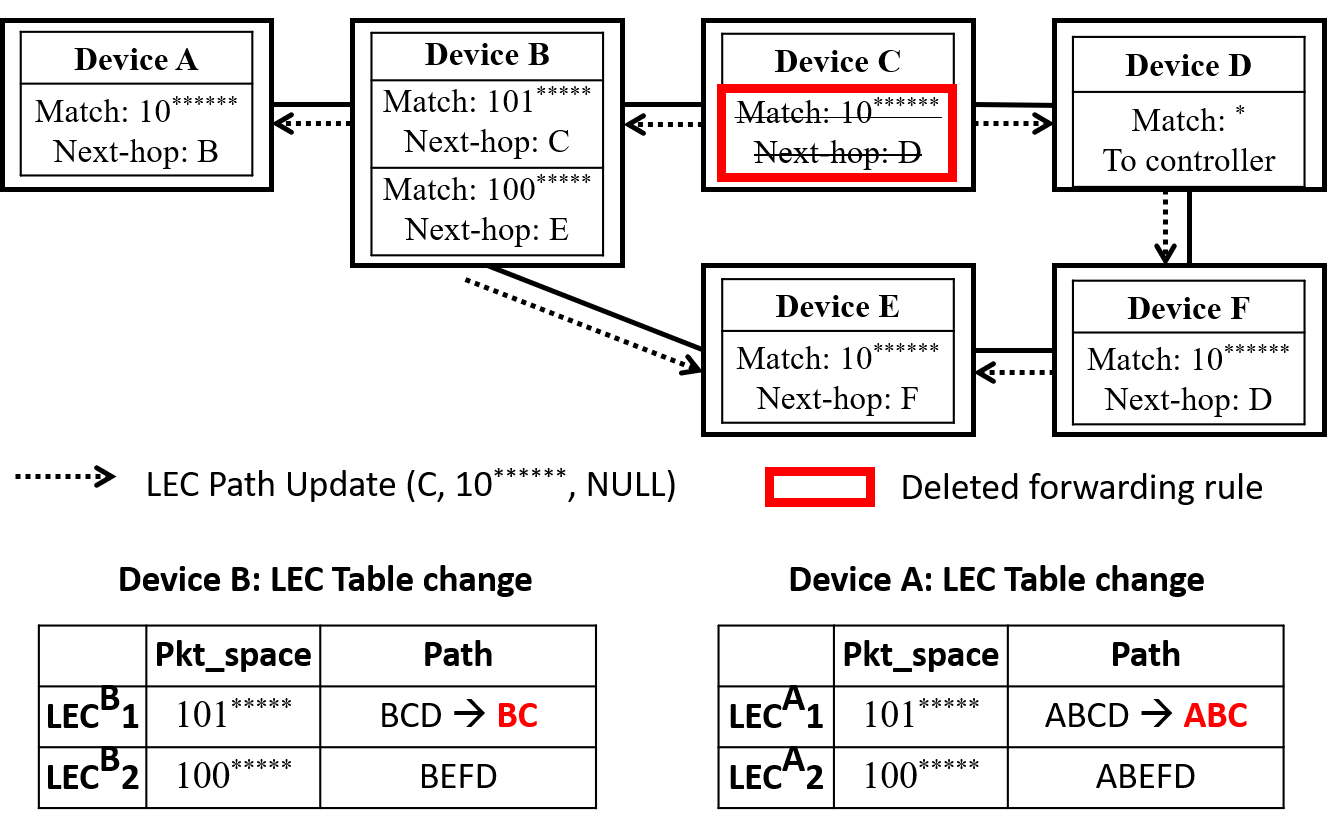} 
    \\ \small(a) & \small (b)
    \\
    \hline
    \includegraphics[width=0.8\columnwidth]{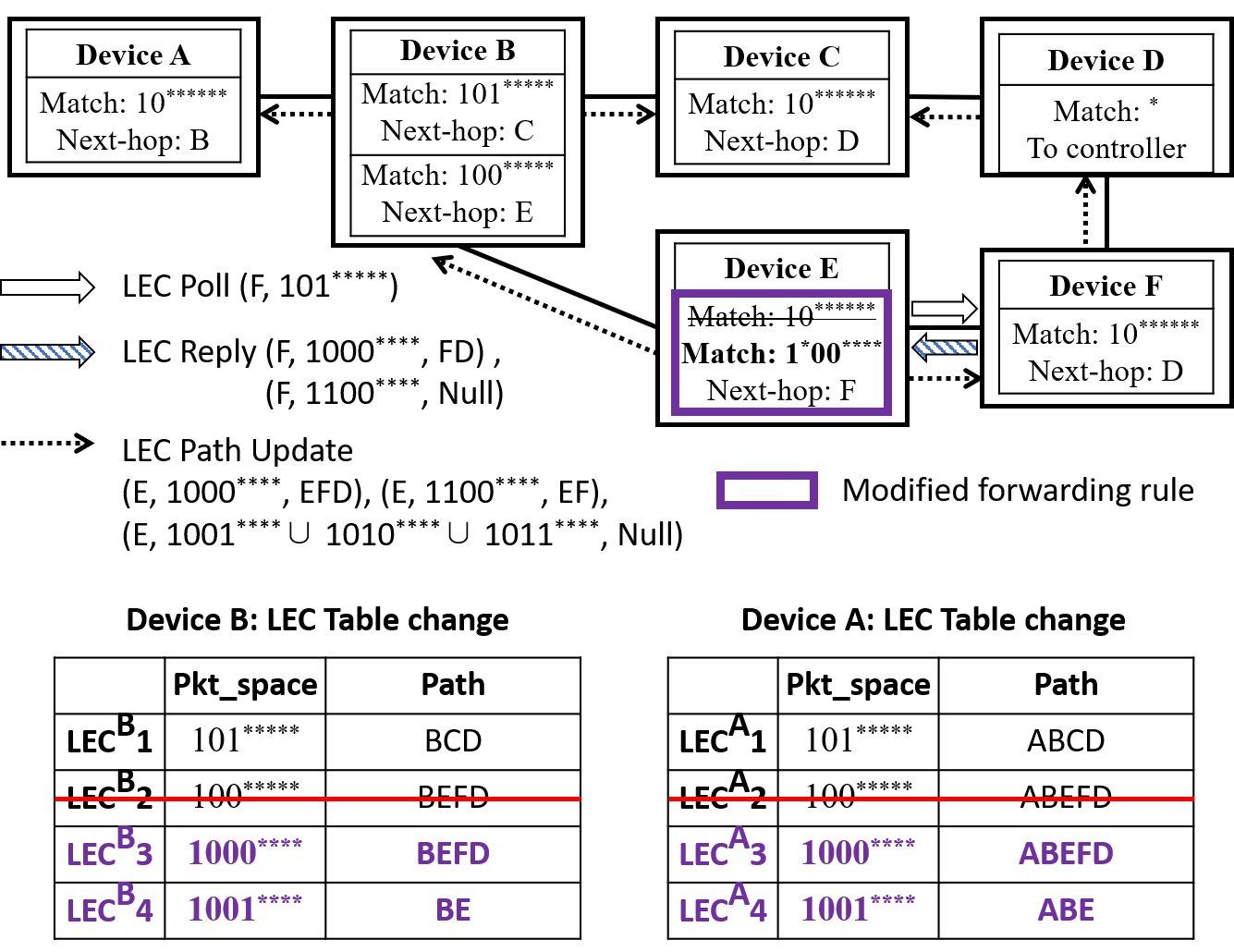} & \includegraphics[width=0.8\columnwidth]{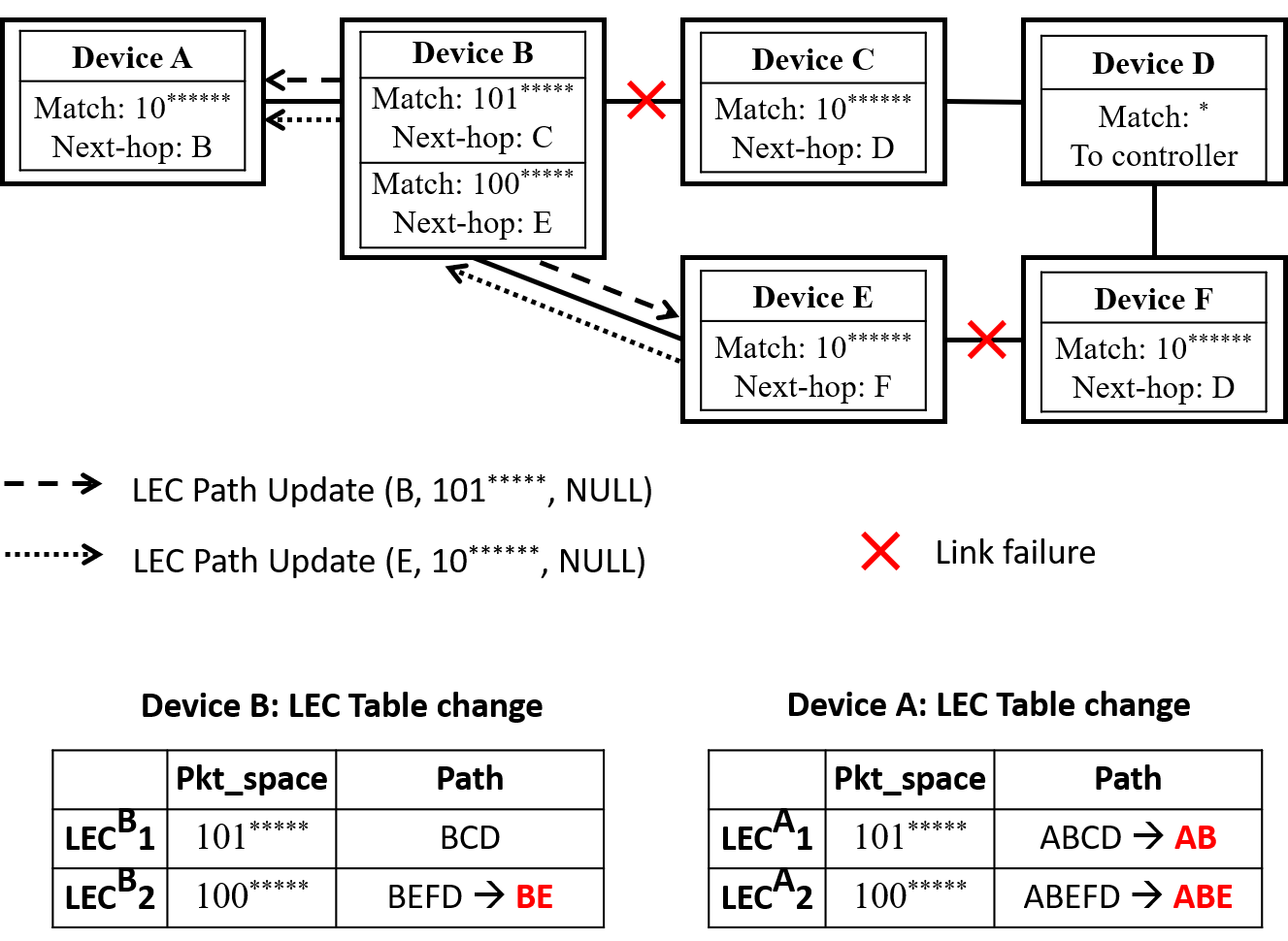} 
    \\ \small(c) & \small (d)
    \\\hline
  \end{tabular}
    \caption{\small{\FSD behavior in response to (a) E1: a forwarding rule is modified at $B$, (b) E2: a deleted forwarding rule at $C$, (c) E3: a forwarding rule is modified at $E$, and (d) E4: a network partition by two link failures.}}\label{dpv-update}
    \reducespace
\end{figure*}

\para{E1: modification of a forwarding rule at $B$.} As shown in Figure \ref{dpv-update}(a), when device $B$ receives its new forwarding rule $<101^{*****}, $ Next-hop: $E>$, $B$ does not have the downstream routing information for $101^{*****}$. As such, $B$ first sends an LEC request to the rule's specified next-hop, \ie, $E$. In response, $E$ returns the reply message for $101^{*****}$, and the local path is $EFD$. Upon receiving the reply message, $B$ calculates the new path segment for $101^{*****}$ and broadcasts the LEC path update $(B, 101^{*****}, BEFD)$ to all devices in the network. Upon receiving this update, $A$ notes that $101^{*****}$ has intersection with its $\mbox{LEC}^A1$ and the device (\ie, $B$) specified in the incoming update message is contained in the local path of $\mbox{LEC}^A1$, so $A$ updates its LEC table with the entry $(101^{*****}, ABEFD)$ based on the new segment. $\mbox{LEC}^A1$ and $\mbox{LEC}^A2$ will further be joined as one LEC. Nevertheless, $C$, $E$ and $F$ do not update their LEC tables because they do not have any path dependency on $B$. Adding a forwarding rule follows the same workflow.

\para{E2: deletion of a forwarding rule at $C$.}
Figure \ref{dpv-update}(b) shows the deletion of a flow rule at $C$. On detecting the flow rule deletion, $C$ updates its local path in the LEC table, and broadcasts a LEC path update $(C, 10^{******}, NULL)$ across the network. Since $A$ and $B$ have dependencies on both $10^{******}$ and $C$, they update the segments after $C$ in their affected LEC entries ($101^{*****}$). In contrast, other devices do not update their LEC entries because although they may have packet space dependencies, they do not have any path dependency.

The first two event examples mainly illustrate the work flow of the distributed verification protocol upon a forwarding state change.
From the example we can see, the protocol can converge quickly because (1) it utilizes a link-state-based update algorithm allowing devices (\eg, $A$ and $B$)  to update their LEC tables in parallel after receiving the broadcast (\eg, from $C$), and (2) given a forwarding rule modification, only one broadcast is needed: \eg, after $C$ broadcasts the changes, even though $B$ updates its LEC table, $B$ \emph{need not} to broadcast its update. As a result, in a single update, network devices (\eg, $A$) will not receive and need to process multiple broadcasts. The third example below will focus on the LEC computation illustration.

\para{E3: modification of a forwarding rule at $E$.} Figure \ref{dpv-update}(c) shows a forwarding rule modification at $E$. Following the same workflow as E1, $E$ first sends request to $F$, but after doing the intersection at $F$, the packet space $1^*00^{****}$ is partitioned into two parts $1000^{****}$ and $1100^{****}$ with different local paths, which are returned to $E$ separately. After receiving them, $E$ creates the two new LECs in its table, and more importantly, generates and broadcasts three update messages: two for the new LECs and one for the complement of them in $10^{******}$, where the last has no local path. Each device then will process with all of the three packet spaces, and updates its LEC table accordingly. For example, $\mbox{LEC}^B2$ at $B$ has path dependecy with $E$, and after intersection, it becomes $\mbox{LEC}^B3$ and $\mbox{LEC}^B4$ with different local paths.

\para{E4: network partition by two link failures at $BC$ and $EF$.} On the event of a network partition, two portions of the network may be completely separated from each other. Furthermore, in the case of an SDN control plane, some devices may be unable to communicate with the SDN controller and therefore be unable to respond to the link failures. \FSD allows these devices to quickly detect which devices are still reachable even in the absence of communication with the control plane. In Figure \ref{dpv-update}(d), device $C$ detects the link failure of $CE$ and notes that this affects both of its forwarding rules. $C$ broadcasts an LEC path update message for the match sets of each forwarding rule, as depicted, and device $A$ updates its LEC table accordingly. Device $A$ also detects the $AB$ link failure, but none of its forwarding rules depend on that link, so it sends no updates. Thus device $A$ can quickly detect its partition from $D$ and $E$ while recognizing reachability to $C$.

\para{Concurrency of multiple network events.} Previous examples only considered network events separately. In reality, multiple updates (\eg, E1 and E2) may  -- and often will -- occur nearly simultaneously due to the controller sending independent updates in parallel, or a distributed CP running. In such instances, our distributed verification protocol will always converge independently of the order of the events.
The protocol guarantees \textbf{eventual consistency} as a link-state protocol because eventually all devices can receive a consistent forwarding state, as long as the CP itself has converged. 
Depending on the order of the updates being processed, unnecessary verification steps may be executed before reaching the final stable state. 
\section{Implementation}\label{sec:implementation}

We develop \impl, a switch OS level verification software suite and an evaluation framework which supports hybrid control planes, to implement \system described in Section \ref{sec:overview}. \impl is mostly written in Python 3.6. \impl is designed in a data driven model by realizing a single in-memory data store to decouple modules. 
A regular switch in \impl is mainly composed of 5 components:

\begin{itemize} [leftmargin=*]
 
    \item \textbf{In-memory datastore.} A key-value datastore (\ie, Redis~\cite{redis}) that stores \emph{local forwarding rules} and \emph{verification results} for all control planes, and exposes \emph{Get/Put/Delete/Listen} APIs for manipulating data and listening events. 
    
    \item \textbf{Verification thread pool.} A thread pool that executes \DistVeri for each control plane. It listens to forwarding rule change events from the \textit{in-memory datastore} to trigger verification updates. 
    We use a thread pool here for 2 main reasons: (1) there might be multiple messages received from neighbors that we need to handle concurrently. Threadpool reduces the overhead for starting a new thread, (2) the thread pool size can be adjusted at any time to control the performance and resource usage of verification (\eg, one could reduce the thread pool size to limit the resource usage). Finally, the verification result will be sent to the \emph{in-memory datastore}.
    
    \item \textbf{Verification result composer.} A component that listens and reads the verification results in \emph{in-memory datastore} and performs the composition of CPs based on the verification results.
    
    \item \textbf{Verification message dispatcher.} A local Openflow controller application built on top of the Ryu framework\cite{ryu} that collects and dispatches messages for the \emph{Verification Thread Pool}. It uses Openflow v1.3 to communicate with the Openflow datapath. Verification messages are encapsulated into IPv4 packets with IP protocol number 143.
    \item \textbf{Consistent update.} We implement the consistent update framework from~\cite{jin2014dynamic} at each node to guarantee strong consistency (\eg, loop-freedom) when applying data plane updates.
\end{itemize}



\begin{figure}
\setlength{\abovecaptionskip}{0.1cm}
\setlength{\belowcaptionskip}{-0.cm}
\includegraphics[width=0.5\columnwidth]{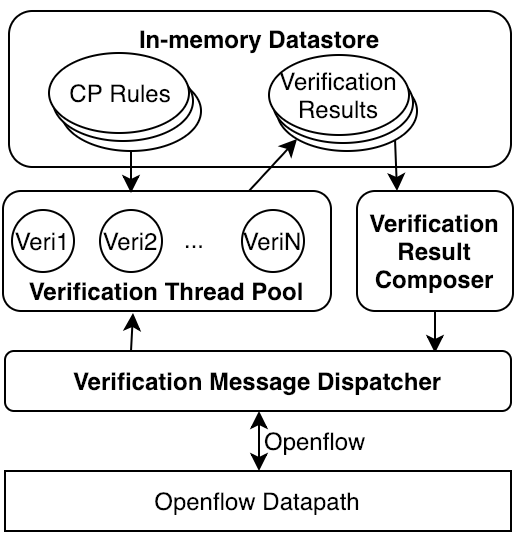}\caption{\small{The Architecture of MultiJet.}}\label{fig:multijet}
\reducespace
\vspace{-0.5em}
\end{figure}

\section{Experiment Settings}\label{sec:eval-set}
We evaluate \system in a virtualized environment + real white-box switches. We run both emulation and simulation on a variety of networks described in Table~\ref{tab:topologies}.
In emulation experiments, each switch is modeled as a separate container using Docker~\cite{docker}, and runs CPs provided by Quagga~\cite{quagga}, OpenFlow-based SDN and \impl. 
We run the Rocketfuel topology on a dedicated server with 2 Intel Xeon 8168 CPU (2.70GHz) having 48 cores and 384 GB memory, and we connect it with two \emph{real Openflow white-box switches} Pica8 P-5401 and Dell Z9100-ON. We run Stanford and AT\&T backbone topologies on a virtual machine with AMD EPYC 7571 (2.1 GHz) CPU having 32 vCPUs and 128 GB memory on Amazon Web Services (AWS)~\cite{aws}.
\end{appendices}
\end{document}